\documentclass[12pt,epsfig,epsf]{article}
\usepackage{amsmath}
\usepackage{cite}
\usepackage{slashed}
\usepackage{epsf,color,colordvi}
\usepackage{graphicx}
\usepackage{amsmath}
\usepackage{amssymb}
\usepackage{enumerate}
\usepackage{epsfig}
\usepackage{cite}
\usepackage{fontenc}
\usepackage{float}
\usepackage[lofdepth,lotdepth,caption=false]{subfig}
\usepackage{xcolor}
\usepackage{booktabs}
\usepackage{tabularx}
\usepackage{multirow}
\usepackage{longtable}
\usepackage{adjustbox}
\usepackage{lscape}
\usepackage{dcolumn}
\usepackage{rotating}
\usepackage[english]{babel}
\usepackage[autostyle]{csquotes}
\usepackage{hyperref}
\usepackage[normalem]{ulem}
\usepackage{color}
\newcommand {\ignore}[1]{}
\definecolor{darkgreen}{cmyk}{1,0,1,0.4}
\definecolor{brown}{cmyk}{0,0.8,1,0.2}
\definecolor{darkred}{cmyk}{0,1,1,0.2}

\def\nue{{\nu_e}}

\def\numu{{\nu_{\mu}}}

\newcommand{\eg}{{\it e.g.}}
\newcommand{\ie}{{\it i.e.}}
\newcommand{\etc}{{\it etc.}}

\newcommand{\beq}{\begin{equation}}
\newcommand{\eeq}{\end{equation}}
\newcommand{\beqa}{\begin{eqnarray}}
\newcommand{\eeqa}{\end{eqnarray}}

\newcommand{\ta}{\theta_{12}}
\newcommand{\tb}{\theta_{13}}
\newcommand{\tc}{\theta_{23}}
\newcommand{\td}{\theta_{14}}
\newcommand{\te}{\theta_{24}}
\newcommand{\tf}{\theta_{34}}
\newcommand{\da}{\delta_{13}}
\newcommand{\db}{\delta_{24}}
\newcommand{\dc}{\delta_{34}}
\newcommand{\ldm}{\Delta m_{31}^2}
\newcommand{\sdm}{\Delta m_{21}^2}
\newcommand{\lldm}{\Delta m_{41}^2}

\newcommand{\mue}{\nu_\mu \rightarrow \nu_e}
\newcommand{\pab}{P(\nu_{\alpha} \rightarrow \nu_{\beta})}

\newcommand{\mumu}{\nu_\mu \rightarrow \nu_\mu}

\newcommand{\chisq}{\Delta\chi^2}

\hypersetup{
  colorlinks,
  citecolor=red,
  linkcolor=blue,
  urlcolor=blue}

\begin{document}

\begin{titlepage}

\renewcommand{\thefootnote}{\alph{footnote}}

\vspace*{1.cm}
\begin{flushright}

\end{flushright}


\renewcommand{\thefootnote}{\fnsymbol{footnote}}
\setcounter{footnote}{0}

{\begin{center}
{\Large  Sterile sector impacting the correlations and degeneracies among mixing parameters at the Deep Underground Neutrino Experiment
\\
}
\end{center}}

\renewcommand{\thefootnote}{\alph{footnote}}

\vspace*{.8cm}
\vspace*{.3cm}
{
\begin{center} 

   {\sf                 Sabila Parveen$^{\S}$\,  \footnote[1]{\makebox[1.cm]{Email:}  sabila41{\_}sps@jnu.ac.in},
                }
        {\sf               Mehedi Masud$^{\P}$\,  \footnote[2]{\makebox[1.cm]{Email:} masud@cau.ac.kr}
        }
            {\sf               Mary Bishai$^{\dag}$\,  \footnote[3]{\makebox[1.cm]{Email:} mbishai@bnl.gov}        
                }                
and                 
            {\sf                 Poonam Mehta$^{\S}$\,\footnote[4]{\makebox[1.cm]{Email:} pm@jnu.ac.in}
}
\end{center}
}
\vspace*{0cm}
{\it 
\begin{center}
$^\S$\, School of Physical Sciences, Jawaharlal Nehru University, 
      New Delhi 110067, India  \\
$^\P$\, High Energy Physics Center, Chung-Ang University, Seoul 06974, Korea    \\
$^\dag$\, Brookhaven National Laboratory, P.O. Box 5000, Upton, NY 11973, USA
\end{center}
}

\vspace*{1.5cm}

{\Large 
\bf
 \begin{center}
  Abstract  
\end{center} 
}

We investigate the physics potential of the upcoming Deep Underground Neutrino Experiment (DUNE) in probing active-sterile mixing.  We present analytic expressions for relevant  oscillation probabilities for three active and one sterile neutrino of eV-scale mass and highlight essential parameters  impacting the oscillation
signals at DUNE. We then explore the space of sterile parameters as well as study their correlations among
themselves and with parameters appearing in the standard framework ($\delta_{13}$ and $\theta_{23}$). 
 We perform a combined fit for the near and far detector at DUNE using GLoBES. We consider alternative beam tune  (low energy and medium energy) and runtime combinations for constraining the sterile parameter space. We show that 
 charged current and neutral current interactions over the near and far detector at DUNE allow for an improved sensitivity for a wide range of sterile neutrino mass splittings. 

\vspace*{.5cm}

\end{titlepage}

\section{Introduction}
\label{sec:intro}

The discovery of neutrino oscillations~\cite{nobel2015} implies that neutrinos are massive and they mix. The three neutrino mixing forms the standard paradigm to explain most of the data from various solar, atmospheric, reactor and  accelerator experiments (for global analysis of oscillation data, see~\cite{10.5281/zenodo.4726908, 
deSalas:2020pgw, nufit_globalfit, Esteban:2020cvm, Capozzi:2018ubv}). According to this, there are  three active flavors of neutrinos $\nu_e, \nu_\mu, \nu_\tau$ which  interact via weak interactions in the Standard Model and are superpositions of the three massive neutrinos $\nu_1,\nu_2,\nu_3$ with masses $m_1,m_2,m_3$ respectively. Neutrino mixing is described by two mass-squared differences, $\ldm$ and $\sdm$, the three mixing angles, $\ta, \tb, \tc$ and the CP phase, $\da$~\cite{Pontecorvo:1957qd,Pontecorvo:1957cp,Gribov:1968kq,Maki:1962mu}. We refer to this as the $(3+0)$ framework.
However, there also exist some anomalies, particularly in short-baseline (SBL) oscillation experiments which cannot be accommodated in standard three-flavor neutrino oscillation framework.  The data hints towards an additional (sterile) neutrino (referred to as the $(3+1)$  framework) with a mass-squared difference, $\lldm \sim \mathcal{O}(1)\text{ eV}^{2}$. The possible existence of sterile neutrinos is one of the most widely explored topics in present day particle physics (see~\cite{Giunti:2019aiy, Dasgupta:2021ies} for reviews).

The $(3+1)$ neutrino framework is well-motivated from data corresponding to three classes of experiments:\begin{enumerate}
\item[(i)] Excess of $\nu_{e}/\bar{\nu}_{e}$-like 
events in accelerator-based SBL oscillation experiments such as Liquid Scintillator Neutrino
Detector (LSND)~\cite{Hill:1995gf,LSND:1996ubh,LSND:1997vun,LSND:2001aii}
 ($\sim 4\,\sigma$ significance)
 and MiniBooNE~\cite{MiniBooNE:2013uba, MiniBooNE:2020pnu, Arguelles:2021meu} ($\sim 5\,\sigma$). {Moreover, the combined significance of LSND and MiniBooNE is around $6.1\,\sigma$\ \cite{MiniBooNE:2020pnu}.}
\item[(ii)] Around $3\,\sigma$ rate deficit of $\bar{\nu}_{e}$-like events in 
 reactor experiments, also known as reactor anti-neutrino anomaly (RAA)~\cite{Mention:2011rk, Mueller:2011nm, Huber:2011wv}. Regarding the reactor experiments, though Neutrino-4 has recently claimed to have observed sterile neutrino (with $\lldm \simeq 7 \text{ eV}^{2}$ and $\sin^{2}\td \simeq 0.09$ at more than $3\,\sigma$ confidence level)~\cite{Serebrov:2020rhy, Serebrov:2020kmd, neutrino4_nu2020}.  {Furthermore, the recent results show that RAA might be the issue of flux calculation and not the neutrino deficit as was thought earlier~\cite{Berryman:2020agd, Kopeikin:2021ugh, Giunti:2021kab}.}
\item[(iii)] A rate deficit of ${\nu}_{e}$ events in {Gallium-based} radio-chemical 
 experiments ($\sim 3\,\sigma$ significance), also known as Gallium anomaly~\cite{GALLEX:1997lja, SAGE:1998fvr, Kaether:2010ag, Giunti:2010zu, Barinov:2021asz}. The gallium anomaly has also been found to be in tension with 
solar and reactor neutrino data~\cite{Giunti:2022btk, Giunti:2023kyo} and 
the significance can reduce with consideration of more precise 
 neutrino cross-sections or other possible effects from nuclear physics~\cite{Berryman:2021yan, Giunti:2022xat, Huber:2022osv, Brdar:2023cms, Elliott:2023cvh}. 
\end{enumerate}

It may also be pointed out that there are some conflicting results challenging the idea of a light sterile neutrino. These are:
\begin{enumerate}
\item[(i)]
Another SBL accelerator experiment, MicroBooNE did not observe any $\nu_{e}$-like event excess~\cite{MicroBooNE:2021zai, MicroBooNE:2021nxr, MicroBooNE:2021tya, MicroBooNE:2021wad, MicroBooNE:2021pvo, MicroBooNE:2022sdp}. It was claimed~\cite{Arguelles:2021meu} that MicroBooNE did not explore the entire sterile parameter space favored by MiniBooNE. 
{However, by using electron neutrino disappearance data, MicroBooNE hints at a preference for $\sin^{2} \theta_{14} = 0.35$ and $\lldm = 1.25~\textrm{eV}^{2}$ at $2.4\,\sigma$~\cite{Denton:2021czb}.}
    \item[(ii)]
    The hint for a light sterile neutrino from the RAA has weakened in light of recent reactor antineutrino flux spectrum analyses~\cite{Giunti:2021kab, Zhang:2023zif}.
    Reactor experiments such as PROSPECT~\cite{PROSPECT:2020sxr}, STEREO~\cite{STEREO:2022nzk}, DANSS~\cite{DANSS:2018fnn}, NEOS~\cite{NEOS:2016wee}, and the combined analysis of NEOS and RENO~\cite{RENO:2020hva} have not confirmed the Neutrino-4 result independently.
    \item[(iii)]
    There exists a tension between global appearance and disappearance data sets, disfavoring the $(3+1)$ scenario at a $4.7\,\sigma$ confidence level~\cite{Dentler:2018sju}. {Incorporating damping effects into the $(3 + 1)$ model for reactor data sets  reduces the tension in the appearance and disappearance data sets~\cite{Diaz:2019fwt, Hardin:2022muu}}.
    \end{enumerate}

There are several next-generation neutrino oscillation experiments in the pipeline for exploring the unresolved 
issues in standard three-flavor neutrino oscillation such as, the search for leptonic CP violation (CPV), determination of neutrino mass hierarchy and the determination of the correct octant of 
the mixing angle $\tc$. 
These future experiments include, for \eg,  Deep Underground Neutrino Experiment (DUNE)~\cite{Acciarri:2015uup, DUNE:2020ypp,DUNE:2021cuw,DUNE:2020fgq}, Tokai to Hyper-Kamiokande (T2HK)~\cite{Hyper-KamiokandeProto-:2015xww}, Tokai to Hyper-Kamiokande with a second detector in Korea (T2HKK)~\cite{Hyper-Kamiokande:2016srs}, European Spallation Source $\nu$ Super Beam (ESS$\nu$SB)~\cite{ESSnuSB:2013dql}. 
The improved precision of these future neutrino facilities allows us to probe subdominant effects arising due to new physics (e.g., sterile neutrinos).

In the present work, we shall focus mainly on DUNE and explore its capability to probe the 
relevant parameter space in $(3+1)$ scenario. 
Phenomenological consequences of a light sterile neutrino in the context of long baseline neutrino experiments 
such as DUNE have been discussed by various authors in~\cite{
Berryman:2015nua, 
Gandhi:2015xza, 
Agarwalla:2016xxa, 
Agarwalla:2016xlg, 
Dutta:2016glq, 
Blennow:2016jkn, 
Rout:2017udo, 
Choubey:2017cba, 
Coloma:2017ptb, 
Gandhi:2017vzo,
Tang:2017khg, 
Choubey:2017ppj, 
Chatla:2018sos,
Choubey:2018kqq,
Ghoshal:2019pab, 
Krasnov:2019kdc, 
Majhi:2019hdj, 
Fiza:2021gvq, 
Ghosh:2021rtn,
Penedo:2022etl,
Denton:2022pxt, 
Chatterjee:2022pqg, 
Parveen:2023ixk,
Kaur:2024jko,
DUNE:2020fgq
}. 
DUNE also has the ability to constrain other sub-dominant new physics scenarios such as non-standard interaction (NSI) etc (see~\cite{Farzan:2017xzy,Proceedings:2019qno}).
DUNE offers distinct advantage as it exploits a wide band beam. 
The standard beam is the low energy (LE) flux (having a peak around $2-3$ GeV and sharply falling at energies $E \gtrsim 4$ GeV) with a total runtime of $13$ years distributed equally between the $\nu$ and $\bar{\nu}$ modes~\cite{DUNE:2020ypp,DUNE:2021cuw}.  Among the viable alternative future options for DUNE, there is a possibility of deploying a 
medium energy (ME) beam based on the NuMI focusing system which offers substantial statistics even at energies $E \gtrsim$ $4$ GeV~\cite{dunefluxes}. 
In order to improve the sensitivity to standard unknowns as well as to new physics at DUNE, studies pertaining to optimization of beamtunes and runtime combinations have been recently carried out~\cite{
Masud:2017bcf, 
Masud:2018pig,
Rout:2020cxi, 
Rout:2020emr,
Siyeon:2024pte}. 

To exploit the full potential of DUNE in probing the sterile neutrino parameter space, we utilize an optimized combination of  LE beam tune along with the ME  beam tune.
The present work goes beyond studies existing in the literature in several aspects.
First, we use two beam tunes, LE and ME and study the role of LE and ME in sensitivity to sterile parameters.
Note that the imprints of sterile neutrino can be seen at both near detector (ND) and far detector (FD). This is sensitive to the value of the $\lldm$. If $\lldm \sim \mathcal{O}(1)\text{ eV}^{2}$, we can see sterile oscillations around $L/E \sim \mathcal{O}(1) \text{km/GeV}$ which corresponds to the location of ND for DUNE. We consider not only charged current (CC) interactions (as done in most studies) but also neutral current (NC) channel which provides clear advantage in probing some of the $(3+1)$ parameters~\cite{Gandhi:2017vzo}. Our aim is to provide a comprehensive sensitivity analysis for the $(3+1)$ scenario at DUNE with optimal beamtune and runtime combinations, utilizing both CC and NC channels as well as information gleaned at FD and ND.

The article is organized as follows. 
We begin with probability level discussion to bring out the key differences in the $(3+0)$ and $(3+1)$ scenarios in Sec.\ \ref{sec:prob}. 
We then describe different beam tunes and the corresponding $\nu_\mu \to \nu_e$ event spectra in Sec.\ \ref{sec:event}. 
This is followed by an outline of the statistical procedure adopted (Sec.\ \ref{sec:chisq}). We discuss the results of our sensitvity analysis in Sec.\ \ref{sec:results}. We summarize and conclude in Sec.\ \ref{sec:conclusion}.

\section{Probability level discussion}
\label{sec:prob}
The oscillation probability in vacuum for any number of flavours (including the sterile ones) can be expressed as 
\beqa \label{eq:pab_gen}
\nonumber \pab &=& \delta_{\alpha\beta}-4Re\sum_{i<j}(U_{\alpha i}U^{*}_{\beta i} U^{*}_{\alpha j}U_{\beta j})\sin^2\Delta_{ji}
+ 2Im\sum_{i<j}(U_{\alpha i}U^{*}_{\beta i}U^{*}_{\alpha j}U_{\beta j})\sin2\Delta_{ji}.
\eeqa

Here $i,j$ run over the mass eigenstates, whereas $\alpha,\beta$ denote flavours. Additionally, 
$\Delta_{ji} = 1.27 \times \Delta m^{2}_{ji}[\textrm{eV}^{2}] \times {L[\textrm km]}/ {E[\textrm {GeV}]}$ where  $L$ is the baseline
length and $E$ is the neutrino energy.
We  shall adopt the following parameterization for the mixing matrix (as in \cite{Gandhi:2015xza, Gandhi:2017vzo}) 
\beq
U_{\text{}}^{(3+1)}=O(\theta_{34},
\delta_{34})O(\theta_{24},\delta_{24})O(\theta_{14})
\underbrace{{O(\theta_{23})O(\theta_{13},
\delta_{13})O(\theta_{12})
}}_{U_{\text{PMNS}}}
\label{eq:U_matrix}\eeq
where $O(\theta_{ij},\delta_{ij})$ is a rotation matrix
in the $ij$ sector with associated phase $\delta_{ij}$ \ie, \beq
O(\theta_{24},\delta_{24}) = 
\begin{pmatrix}
1 & 0 & 0 & 0 \\
0 & \cos\tb & 0 & e^{-i\delta_{24}}\sin\tb \\
0 & 0 & 1 & 0 \\
0 & -e^{i\delta_{24}}\sin\tb & 0 & \cos\tb
\end{pmatrix};~
O(\theta_{12}) = 
\begin{pmatrix}
\cos\ta & 0 & 0 & \sin\ta \\
0 & 1 & 0 & 0 \\
0 & 0 & 1 & 0 \\
-\sin\ta & 0 & 0 & \cos\ta
\end{pmatrix} \text{\etc}
\eeq
A close examination of  Eq.\ \eqref{eq:U_matrix} reveals that $\theta_{23}$  and  $\delta_{34}$ do not appear in the first row and last column of the mixing matrix. 
As a result, it is possible to extract five of the six mixing angles using this parametrization\ \cite{Gandhi:2015xza}. Some other forms are also used in the literature\ \cite{Kopp:2013vaa, Klop:2014ima, DUNE:2020fgq}.

Since we are interested in distinguishing between standard $(3+0)$ and sterile $(3+1)$ cases, let us define a quantity, 
$\Delta P_{\alpha \beta} = P^{(3+0)}_{\alpha \beta} - P^{(3+1)}_{\alpha \beta}$,
where $P^{(3+0)}_{\alpha \beta}$ is the probability for the $(3+0)$ case and $P^{(3+1)}_{\alpha \beta}$ is the probability for $(3+1)$ case.  For the probability expressions for the $(3+0)$ case and the $(3+1)$ case in vacuum\,\footnote{The probability expressions in the two cases have been computed in matter as well\ \cite{Klop:2014ima,:2019wbn}. For our purpose, it suffices to understand the general features of the probabilities using the vacuum expressions. 
}, 
we refer the reader to Appendix~\ref{appendix_a}. 
Here, we present the approximate expressions for $ \Delta P_{\alpha \beta}$ in vacuum corresponding to the three channels relevant to our analysis, namely the $\mue$ channel\ \cite{Gandhi:2015xza}, the $\mumu$ channel and the NC channel\ \cite{Gandhi:2017vzo}. The expressions listed here are valid at the FD location of DUNE ($L = 1300$ km). In writing these expressions, we have used certain approximations - 
\begin{enumerate}
\item[(a)]   Neglecting the smallest mass-squared difference, $\sdm$ and $\Delta m^2_{31} \simeq \Delta m^2_{32}$ 
\item[(b)] $\Delta m^2_{41} \simeq \Delta m^2_{42} \simeq \Delta m^2_{43} $   
\item[(c)] Averaging of oscillations induced by the largest mass-squared difference,  $\lldm \sim 1 \text{ eV}^{2}$  
\end{enumerate}
Further, $\db$ and $\dc$ are taken to be zero and $\tc = \pi/4$. Since $\theta_{13} , \theta_{14} , 
\theta_{24} \leq 13^\circ$~\cite{Dentler:2018sju},  we have set $\sin\theta_{ij} \sim 10^{-1} \sim \lambda$, where $
\theta_{ij}$ could be $\theta_{13}$, $\theta_{14}$ or $\theta_{24}$.

\begin{align}
\label{eq:delta_pmue}
\Delta P_{{\mu} e}(\theta_{14}, \theta_{24})  &\simeq 
\dfrac{1}{2} \sin^{2} 2\theta_{13} \Big[1 - \cos^{2} \theta_{14} \cos^{2} \theta_{24} \Big] 
\sin^2\dfrac{\Delta m^2_{31}L}{4E} \nonumber \\
&- \frac{1}{\sqrt{2}}
\sin2\td \sin\te \cos\td \cos\te \sin 2\tb 
\sin \Big(\frac{\Delta m^2_{31}L}{4E} + \da \Big) 
\sin \frac{\Delta m^2_{31}L}{4E} \nonumber \\
&-\frac{1}{2}\sin^{2}2\td \sin^{2}\te 
+ \mathcal{O}(\lambda^{5}),
\\[3mm]
\label{eq:delta_pmumu}
 \Delta P_{\rm{\mu}\mu} (\theta_{14}, \theta_{24}) &\simeq 
 2 \cos^{2}\theta_{14} \sin^{2} \theta_{24} - \cos^{2}\theta_{13}\sin^{2}\theta_{24}\sin^2\dfrac{\Delta m^2_{31}L}{4E}  \nonumber \\
&- \cos^{2}\theta_{13}\sin^{2}\theta_{24}
 \Big[\sin^{2} \theta_{13} - \sin^{2} \theta_{24} 
 - 2\sin^{2}\theta_{14}\cos^{2}\theta_{24} \Big] \sin^2\dfrac{\Delta m^2_{31}L}{4E} \nonumber \\
 &+ \mathcal{O}(\lambda^{5}),
\\[3mm]
\label{eq:delta_pnc}
\Delta P_{{NC}}^{\text{}} (\theta_{14}, \theta_{24}, \theta_{34}) &\simeq 
\frac{1}{2} \cos^{4}\theta_{14} \cos^{2}\theta_{34} \sin^{2} 2\theta_{24} 
+ \cos\theta_{13}\cos^{2}\theta_{24}
\Big[\cos^3\theta_{13} \sin^2\theta_{34} 
\nonumber \\
&-\cos\theta_{13} \cos^2\theta_{34} \sin^2\theta_{24} 
+ \sqrt{2} \sin\theta_{13} \sin2\theta_{34}\sin\theta_{14} \cos\theta_{24} \Big] \sin^2\dfrac{\Delta m^2_{31}L}{4E} \nonumber \\
&+ \mathcal{O}(\lambda^{4}).
\end{align}
The parentheses on the left hand side (LHS) of Eq.\ \eqref{eq:delta_pmue}, Eq.\ \eqref{eq:delta_pmumu} and Eq.\ \eqref{eq:delta_pnc} highlight which ones of the active-sterile mixing angles ($\theta_{14} , \theta_{24}, \theta_{34}$) impact the particular channel.
 \begin{table}[h]
\centering
\scalebox{0.9}{
\begin{tabular}{| c | c | c | c |}
\hline
&&&\\
Parameter & Best-fit-value & $3\,\sigma$ interval & $1\,\sigma$ uncertainty  \\
&&&\\
\hline
&&&\\
$\theta_{12}$ [Deg.]             & 34.3                    &  31.4 - 37.4   &  2.9\% \\
$\theta_{13}$ (NH) [Deg.]    & 8.53  &  8.13  -  8.92   &  1.5\% \\
$\theta_{13}$ (IH) [Deg.]    & 8.58  &  8.17  -  8.96   &  1.5\% \\
$\theta_{23}$ (NH) [Deg.]        & 49.3      &  41.2  - 51.4    &  3.5\% \\
$\theta_{23}$ (IH) [Deg.]        & 49.5                     &  41.1  - 51.2    &  3.5\% \\
$\sdm$ [$\text{eV}^2$]  & $7.5 \times 10^{-5}$  &  [6.94 - 8.14]$\times 10^{-5}$  &  2.7\% \\
$\ldm$ (NH) [$\text{eV}^2$] & $2.55 \times 10^{-3}$   &  [2.47 - 2.63] $\times 10^{-3}$ &  1.2\% \\
$\ldm$ (IH) [$\text{eV}^2$] & $-2.45 \times 10^{-3}$  & $-[2.37$ - $2.53]\times 10^{-3}$  &  1.2\% \\
$\delta_{13}$ (NH) [Rad.]   & $-0.92\pi$   & $[-\pi, -0.01\pi]  \cup [0.71\pi, \pi]$ &  $-$ \\
$\delta_{13}$ (IH) [Rad.]   & $-0.42\pi$   & $[-0.89\pi, -0.04\pi]$   & $-$  \\
$\lldm$ [$\text{eV}^2$]   & $1$   & $-$   & $-$  \\
$\td$ [Deg.] & 5.7 & 0 - 18.4 & $\sigma(\sin^{2}\theta_{14}) = 5\%$ \\
$\te$ [Deg.] & 5 & 0 - 6.05 & $\sigma(\sin^{2}\theta_{24}) = 5\%$ \\
$\tf$ [Deg.] & 20 & 0 - 25.8 & $\sigma(\sin^{2}\theta_{34}) = 5\%$ \\
$\delta_{24}$ [Rad.]   & $0$   & $[-\pi, \pi]$ &  $-$ \\
$\delta_{34}$ [Rad.]   & $0$   & $[-\pi, \pi]$ &  $-$ \\
&&&\\
\hline
\end{tabular}}
\caption{\footnotesize{\label{tab:parameters}
 The values of the oscillation parameters and their uncertainties used in our study. 
 The values of standard (3+0) parameters were taken from the global fit analysis in \cite{10.5281/zenodo.4726908, deSalas:2020pgw} while the sterile (3+1) parameter values were chosen from \cite{Dentler:2018sju}.  
If the $3\,\sigma$ upper and lower limit of a parameter is $x_{u}$ and $x_{l}$ respectively, the $1\,\sigma$  uncertainty is $(x_{u}-x_{l})/3(x_{u}+x_{l})\%$~\cite{DUNE:2020ypp}. 
For the active-sterile mixing angles, a conservative $5\%$ uncertainty was used on $\sin^{2}\theta_{i4}$ (i = 1, 2, 3).
}}
\end{table}
 We have retained terms upto $\mathcal{O}(\lambda^{4})$ in Eq.\ \eqref{eq:delta_pmue} and Eq.\ \eqref{eq:delta_pmumu}, and
upto $\mathcal{O}(\lambda^{3})$ in Eq.\  \eqref{eq:delta_pnc}~\footnote{Note that this approximation is slightly relaxed in comparison to  \cite{Gandhi:2017vzo}.}. 
It may be noted that $\Delta P_{{NC}}^{\text{}} = P_{\mu s}^{\text{}}$ (since $P_{\mu s} ^{(3+0)} = 1$) where $s$ stands for the sterile neutrino (see Eq.\ \eqref{eq:delta_pnc}).
\begin{figure}[t!]
\centering
\includegraphics[scale=2.5]{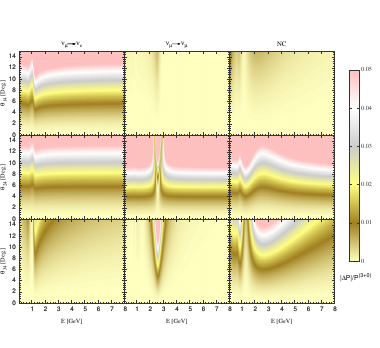}
\caption{\footnotesize{This figure shows the heatplot calculated at the FD with $L = 1300$ km, for the fractional probability difference 
$|\Delta P_{\alpha \beta}|/P_{\alpha \beta}^{(3+0)}$ in matter when neutrino energy $E$ and one of the active-sterile mixing angle is varied individually. 
The three columns illustrate the $\mue, \mumu$ and $NC$ channels respectively. 
The three rows shows the impact of $\td, \te, \tf$ respectively. 
}}
\label{fig:delta_p}
\end{figure}

For better understanding of the impact of sterile sector on the oscillation probabilities, we present heatplots of the fractional probability difference, $|\Delta P_{\alpha \beta}|/P^{(3+0)}_{\alpha \beta}$ (see Appendix\ \ref{appendix_a} and Eq.\ \eqref{p_mue_std}, Eq.\ \eqref{p_mumu_std} and Eq.\ \eqref{p_nc_std} for expressions of $P_{\alpha\beta}^{(3+0)}$)  in the plane of energy, $E$ and $\theta_{ij}$ (with $i=1,2,3$ and $j=4$) for the considered channels in Fig.~\ref{fig:delta_p}. 
The three columns correspond to the three channels ($\mue, \mumu$ and NC), while the three rows depict the dependence on the three active-sterile mixing angles ($\td, \te, \tf$).
 The heatplots are obtained using General Long Baseline Experiment Simulator (GLoBES)~\cite{Huber:2004ka,Huber:2007ji} using the approximation of constant matter density~\footnote{This is a fair approximation for baselines $\leq$ 3000 km~\cite{Gandhi:2004bj}.}.

From Fig.~\ref{fig:delta_p}, for the $\mue$ channel, we note that the dependence on $\td$ (first row, first column) and $\te$ (second row, first column)  is  similar. This can be understood by examining the analytic behaviour of $\Delta P_{\mu e}/P^{(3+0)}_{\mu e}$ taking one parameter at a time,
\begin{align}
\label{eq:delta_pmue_th14}
\frac{\Delta P_{{\mu} e}(\theta_{14})}{P_{{\mu} e}^{(3+0)}}   &\simeq 
\frac{\displaystyle{\lim_{\te \to 0}}
\Big[\Delta P_{{\mu} e} (\td,\theta_{24})\Big]}
{P_{{\mu} e}^{(3+0)}} 
\simeq
\sin^{2}\td , \\
\label{eq:delta_pmue_th24}
\frac{\Delta P_{{\mu} e}(\theta_{24})}{P_{{\mu} e}^{(3+0)}}   &\simeq 
\frac{\displaystyle{\lim_{\td \to 0}}
\Big[\Delta P_{{\mu} e} (\td,\theta_{24})\Big]}
{P_{{\mu} e}^{(3+0)}}
\simeq 
\sin^{2}\te .
\end{align}
We  note that Eq.\ \eqref{eq:delta_pmue_th14} and Eq.\ \eqref{eq:delta_pmue_th24}  depend on $\td$ $(\propto \sin^{2}{\td})$ and $\te$ $(\propto \sin^{2}{\te})$ in a similar manner, which leads to similar dependence on these two parameters. As a result, $|\Delta P_{\mu e}|/P^{(3+0)}_{\mu e}$ grows as $\td$ (or $\te$) increases. 
Note that $\tf$ does not appear in leading order in the expression for $\Delta P_{\mu e}$ (Eq.\ \eqref{eq:delta_pmue}) in vacuum, and the mild variations (kink around $E \simeq 0.9-1.1$ GeV in $\nu_\mu \to \nu_{e}$ channel for both $\td$ and $\te$) seen in Fig.\ \ref{fig:delta_p} can be attributed to the matter effects~\cite{Gandhi:2015xza}. 

For the $\mumu$ channel (second column of Fig.~\ref{fig:delta_p}), there is almost no dependence on $\td$ which follows from Eq.~\eqref{eq:delta_pmumu}. However, this channel depends on $\te$ (second column, second row).
To understand dependence of $\Delta P_{\mu\mu}$ on $\te$, let us examine Eq.~\eqref{eq:delta_pmumu}
  \begin{align}
 \label{eq:delta_pmumu_th24}
 \frac{\Delta P_{\mu\mu}(\te)}
 {P_{\mu\mu}^{(3+0)}}
 &\simeq 
 \frac{\displaystyle{\lim_{\td \to 0}}
 \Big[\Delta P_{\mu\mu}^{\text{}}(\td, \te)\Big]}
 {P_{\mu\mu}^{(3+0)}}\,, \nonumber \\
 &\simeq 
 \frac{2\sin^2\te - \cos^2\tb \sin^2\te \Big[1 -\sin^2\tb - \sin^2\te \Big]  \sin^{2}\Big(\frac{\ldm L}{4E}\Big)
 }
 {1-\Big(\cos^{2}\tb+\frac{1}{4}\sin^{2}2\tb \Big)\sin^{2}\Big(\frac{\ldm L}{4E}\Big)}. 
 \end{align}
There are two terms in the numerator.
The first term is $\propto2\sin^2\te$.
The second term with overall negative sign depends on the mixing between  active and sterile sector. 
When $\sin^{2}(\ldm L/4E)\to1$, the denominator becomes very small (as $\tb$ is small, the coefficient of the $\sin^{2} (\ldm L/4E)$ term in the denominator $\simeq 1$), leading to a kink  at $E=2.5$ GeV.
As regards $\tf$, $\Delta P_{\mu\mu}/P^{(3+0)}_{\mu\mu}$ remains unchanged except in a narrow region around $E \simeq 2.5$ GeV.

For the NC channel (third column of Fig.~\ref{fig:delta_p}), there is almost no dependence on $\td$ (third column, first row) which follows from Eq.~\eqref{eq:delta_pnc} $(\Delta P_{{NC}}(\td)$ is vanishingly small to the leading order). However, this channel depends on $\te$ (third column, second row) and $\tf$  (third column, third row). 
Using Eq.\ \eqref{eq:delta_pnc}, we obtain 
\begin{align}
\label{eq:delta_pnc_th24}
\frac{\Delta P_{{NC}}(\te)}
{P_{{NC}}^{{(3+0)}}}&=
\Delta P_{NC}(\te)
\simeq 
 \displaystyle{\lim_{\substack{\td \to 0 \\ \tf \to 0}}}
 \Big[\Delta P_{NC}(\td, \te, \tf)\Big]\,,
\nonumber \\[3mm]
 &\simeq 
\frac{1}{2} \sin^{2} 2\theta_{24} 
- \frac{1}{4}\cos^{2}\theta_{13} \sin^22\theta_{24} \sin^2 \Big(\frac{\Delta m^2_{31}L}{4E}\Big).
\end{align}
where $P_{NC}^{{(3+0)}} = 1$. The first term in Eq.\ \eqref{eq:delta_pnc_th24} increases $\Delta P_{{NC}}$ irrespective of energy, while the second term leads to an opposite effect. The largest effect is around $E \simeq 2.5$ GeV 
{{for a given value of $\te$ as we can see from Fig.\ \ref{fig:delta_p}  (third column, second row).}}
Using Eq.\ \eqref{eq:delta_pnc}, we also obtain 
\begin{align}
\label{eq:delta_pnc_th34}
\frac{\Delta P_{{NC}}(\tf)}
{P_{{NC}}^{{(3+0)}}}
&= \Delta P_{NC}(\tf)
\simeq\displaystyle{\lim_{\substack{\td \to 0 \\ \te \to 0}}}
 \Big[\Delta P_{NC}(\td, \te, \tf)\Big]\,,
 \nonumber\\[3mm]
& \simeq 
\cos^{4}\theta_{13} \sin^2 \theta_{34} \sin^2 \Big(\frac{\Delta m^2_{31}L}{4E}\Big).
\end{align}
%
{{Thus, the corresponding heatplot  (third column, third row in Fig.\ \ref{fig:delta_p}) shows largest change around $E \simeq 2.5$ GeV for a given value of $\tf$.}}

%
\section{Neutrino beam tunes and $\nu_\mu \to \nu_e$ ($\bar \nu_\mu \to \bar \nu_e$) event spectra at DUNE }  
\label{sec:event}
\begin{figure}[t!]
\begin{centering}
\includegraphics[scale=1]{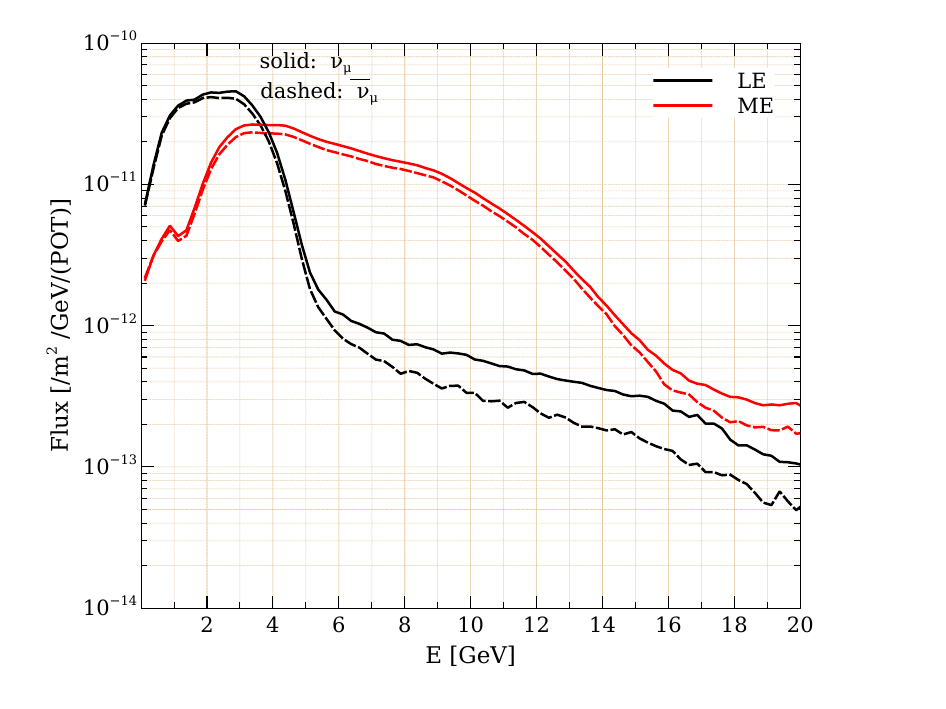}
\caption{Comparison of different beam tunes - black curves represent the LE beam and red curves represent the $\nu_{\tau}$- optimized ME beam as given in DUNE TDR~\cite{DUNE:2020ypp}. 
The solid and dashed curves indicate the $\nu_{\mu}$ and $\bar{\nu}_{\mu}$ flux respectively.}
\label{fig:flux_tdr}
\end{centering}
\end{figure}
As mentioned above, we are interested in differentiating between the  $(3+0)$ and $(3+1)$ cases. For generating the events in the two cases,  we carry out simulations using GLoBES. 
The most recent configuration files from the Technical Design Report (TDR) of DUNE~\cite{DUNE:2020ypp,DUNE:2021cuw} have been used in our simulations. DUNE consists of an on-axis $40$ kiloton (kt) liquid argon FD housed at the Homestake Mine in South Dakota over a baseline of $1300$ km. 
A near detector (ND)  with target mass $0.067$ kt will be installed at a  baseline $0.570$ km at the Fermi National Accelerator Laboratory (FNAL), in Batavia, Illinois. 
 We use the following broad-band beam tunes : 
\begin{enumerate}
\item[(i)] The low energy (LE) beam tune used in DUNE TDR~\cite{DUNE:2020ypp}
\item
[(ii)] The medium energy (ME) beam tune optimized for $\nu_{\tau}$ appearance~\cite{DUNE:2020ypp, dunefluxes}
\end{enumerate}
Both the beams are produced by a $120$ GeV proton beam impinging on a graphite target and are obtained from G4LBNF, a GEANT4 based simulation~\cite{Agostinelli:2002hh,Allison:2006ve} of the long baseline neutrino facility (LBNF) beamline~\cite{DUNE:2020ypp}. 
The hadrons produced in the graphite target are then focussed using three magnetic horns operated with 
$300$ kA current and are allowed to decay in a helium-filled decay pipe of length 194 m to produce the LE flux. 
The higher energy tuned ME flux is simulated by 
replacing the three magnetic horns with two NuMI-like parabolic horns with the second horn starting $17.5$ m downstream from the start of the first horn. 
For both the fluxes, the focusing horns can be operated in forward and reverse current configurations to produce $\nu$ 
and $\bar{\nu}$ beams respectively. 
These two broad-band beam tunes are consistent with what could be technically achieved by the LBNF facility.   
The two beam tunes used in our study are shown in Fig.\ \ref{fig:flux_tdr}. From Fig.\ \ref{fig:flux_tdr} we note that, LE flux  peaks around $\sim 2-3$ GeV and falls rapidly thereafter. 
The ME flux on the other hand peaks around $\sim 3-5$ GeV and falls much slowly thereafter while retaining 
significantly higher flux than LE beam at $E \gtrsim 4$ GeV.
%
\begin{figure}[t!]
\centering
\includegraphics[scale=0.8]{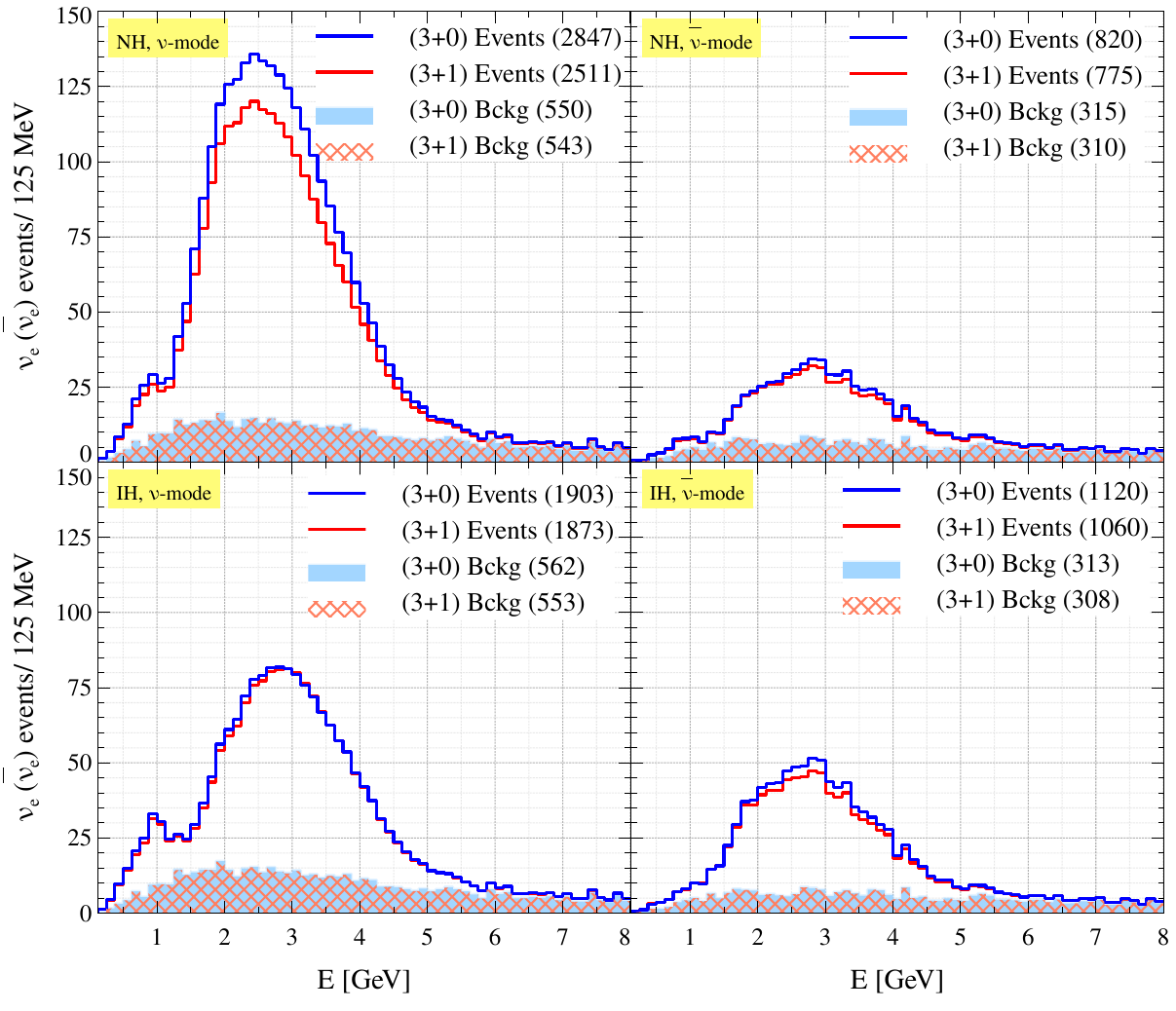}
\caption{The $\nu_{\mu}$ $\to$ $\nu_{e}$ ($\bar{\nu}_{\mu}$ $\to$ $\bar{\nu}_{e}$) event spectra at DUNE FD using the LE beam are shown for the different cases - the top (bottom) row represents the case of NH (IH) while the left (right) column corresponds to $\nu$ ($\bar{\nu}$) mode. 
The total backgrounds corresponding to $(3+0)$ and $(3+1)$ cases are shown as light blue shaded region and light red hatched region respectively. 
A total runtime of $13$ years with $6.5$ years in $\nu$-mode (left column) and $6.5$ years in  
$\bar{\nu}$-mode (right column) has been used to generate the spectra.
The numbers in the legends are the corresponding total number of events and backgrounds summed over the energy bins upto $8$ GeV.}
\label{fig:event_LE}
\end{figure}
 \begin{figure}[t!]
 \centering
\includegraphics[scale=0.8]{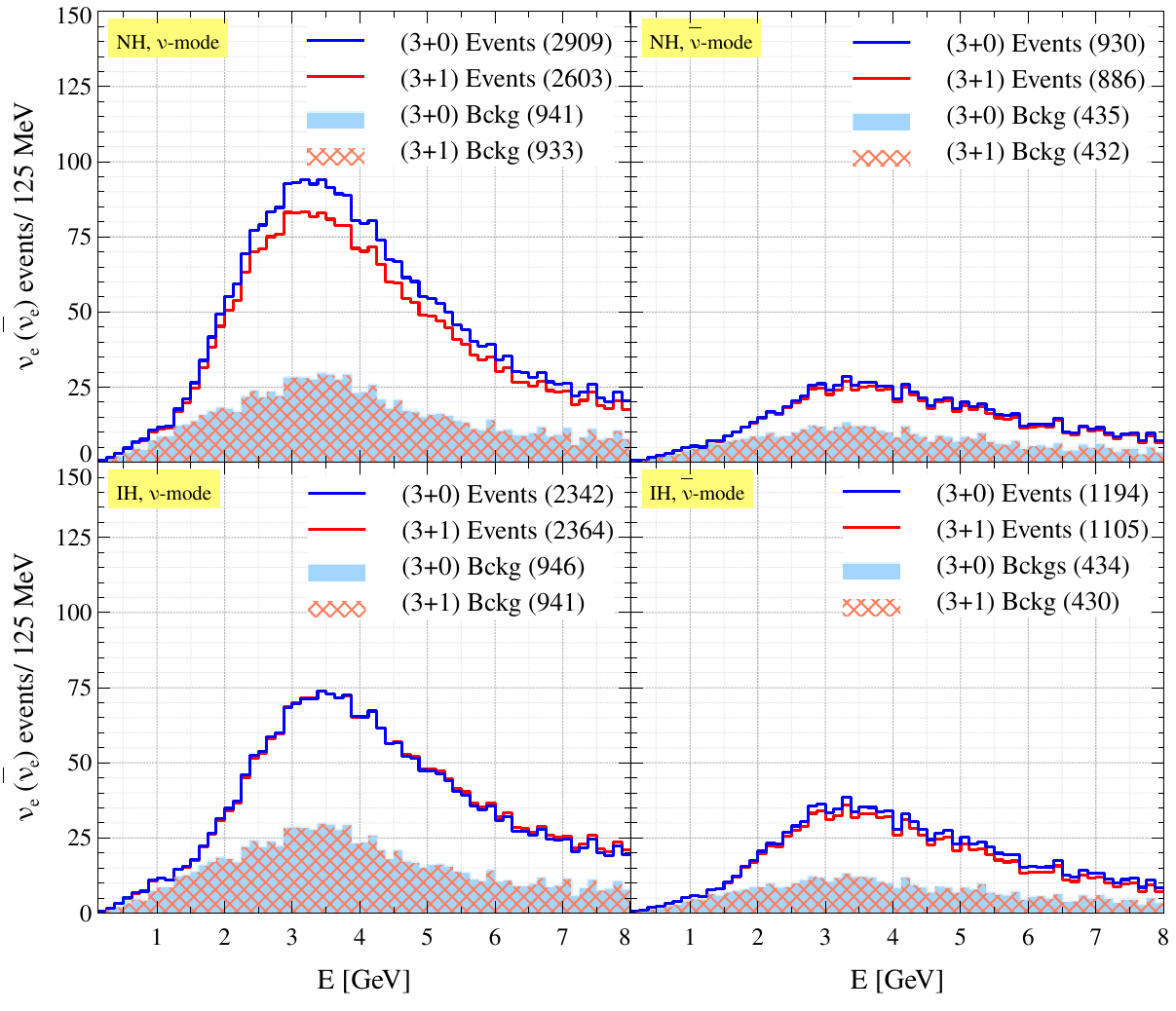}
\caption{Same as Fig.\ \ref{fig:event_LE} but using the $\nu_{\tau}$-optimized ME beam at DUNE FD.}
\label{fig:event_ME}
\end{figure}
In what follows, we discuss the event spectra $(\nu_{\mu}$ $\to$ $\nu_{e}$ ($\bar{\nu}_{\mu}$ $\to$ $\bar{\nu}_{e}))$ obtained at DUNE FD using the two beam tunes given below :
\begin{itemize}
\item LE ($13$), FD, CC 
\item ME ($13$), FD, CC
\end{itemize}
In Fig.\ \ref{fig:event_LE}, we show the $\nu_{\mu}$ $\to$ $\nu_{e}$ ($\bar{\nu}_{\mu}$ $\to$ $\bar{\nu}_{e}$) event spectra at DUNE FD in $(3+0)$ (blue) and $(3+1)$ (red) cases  by using the  LE beam. 
A total runtime of $13$ years with $6.5$ years in $\nu$-mode (left column) and $6.5$ years in  
$\bar{\nu}$-mode (right column) has been used to generate the event spectra. 
For the chosen values of the parameters, the total number of events summed over the energy bins upto $8$ GeV (as written in the legend of Fig.\ \ref{fig:event_LE}) is larger in $(3+0)$ case in comparison to $(3+1)$ case both for $\nu$-mode and $\bar{\nu}$-mode irrespective of the mass hierarchy (if $|\Delta m^2_{31}| > 0$, we have normal hierarchy (NH) or if $|\Delta m^2_{31}| < 0$, we have  inverted hierarchy (IH)).
It may be noted that the highest statistics is attained for the NH, $\nu$-mode as seen in the Fig.\ \ref{fig:event_LE} (top, left panel). 
We also note that the event spectra peaks around $2.5-3$ GeV 
and falls off rapidly beyond  $\sim3$ GeV.
It can be seen that the event spectra corresponding 
to $(3+0)$ and $(3+1)$ case are well separated for NH-$\nu$ mode, with a total of $2847$ events for $(3+0)$ case and $2511$ events for the $(3+1)$ case (smaller by  $12\%$  in comparison to the $(3+0)$ case).
In contrast, the total statistics the total number of events for $(3+0)$ and $(3+1)$ cases differ by a much smaller margin for the other scenarios as shown in the three other panels of Fig.\ \ref{fig:event_LE} (approximately within $1.5-5.5\%$). 
The backgrounds mainly consist of beam impurities from intrinsic $\nu_{e}/
\bar{\nu}$, flavour misidentification, NC and are 
shown as the light blue shaded regions for $(3+0)$ and light red 
hatched regions for $(3+1)$ case. 
The backgrounds summed over the energy bins for these two cases are close to each other are independent of hierarchy. 
The backgrounds for the $\bar{\nu}$-mode are considerably lower than those for the $\nu$-mode. 

In Fig.\ \ref{fig:event_ME}, we show the $\nu_{\mu}$ $\to$ $\nu_{e}$ ($\bar{\nu}_{\mu}\to\bar{\nu}_{e}$) event spectra in $(3+0)$ and $(3+1)$ cases at DUNE FD by using the $\nu_{\tau}$-optimized ME beam.  It may be noted that the event specta peaks around $(3.5-4)$ GeV and fall off slowly thereafter, thus providing considerable statistics at higher energies upto $8$. 
In the analysis that follows, we combine the LE and ME beam tunes  in order to probe the sterile neutrino parameter space, thereby going beyond the analysis using the LE beam tune alone.
\section{Numerical procedure}
\label{sec:chisq}

We explore the capability offered by DUNE to probe the sterile neutrino parameter space by performing a
 $\chisq$ analysis. Even though, we have used GLoBES~\cite{Huber:2004ka, Huber:2007ji} for the $\chisq$ analysis, we can understand the main features by examining the analytic form of the $\chisq$ given below :
\small{
\begin{align}
\label{eq:chisq}
\Delta \chi^{2}(\bar{p}^{\text{fit}})  &= \underset{(p^{\text{fit}}-\bar{p}^{\text{fit}}; \eta)}{{\text{min}}} 
\Bigg[
\color{black}
\underbrace{\color{black}
2\sum_{x}^{\text{mode}}\sum_{j}^{\text{channel}}\sum_{i}^{\text{bin}}
\Bigg\{
N_{ijxy}^{(3+1)}(p^{\text{fit}};\eta) - N_{ijxy}^{(3+0)}(p^{\text{data}}) 
+ N_{ijxy}^{(3+0)}(p^{\text{data}}) \ln\frac{N_{ijxy}^{(3+0)}(p^{\text{data}})}{N_{ijxy}^{(3+1)}(p^{\text{fit}};\eta)} 
}_{\text{statistical}}
\Bigg\} 
\color{black}
\nonumber \\
&+ 
\color{black}
\underbrace{\color{black}
\sum_{l}\frac{(p^{\text{data}}_{l}-p^{\text{fit}}_{l})^{2}}{\sigma_{p_{l}}^{2}}
}_{\text{prior}}
\color{black}
+ 
\color{black}
\underbrace{\color{black}
\sum_{k}\frac{\eta_{k}^{2}}{\sigma_{k}^{2}}
}_{\text{systematics}}
\color{black}
\Bigg],
\end{align}
}
where 
the index $i$ is summed over the energy bins in the range $0.5-10$ GeV\,\footnote{In the present analysis, we have a total of $62$ energy bins in the range $0.5-10$ GeV: $62$ bins each having a width of 
$0.125$ GeV in the energy range of $0.5 - 8$ GeV and $2$ bins of width 1 GeV each in the range 8-10 GeV~\cite{DUNE:2021cuw}.}. 
We have turned on the low-pass filter option in GLoBES with a filter value of 125 MeV in order to smoothen the fast oscillations in underlying probability calculations.  
The index $j$ corresponds to channels ($\nue, \numu$) while the index $x$ runs over the modes ($\nu$ and $\bar{\nu}$).
$N^{(3+0)}$ (treated as \textit{data}) and $N^{(3+1)}$ (treated as \textit{fit}) are the set of events corresponding to the $(3+0)$ and $(3+1)$ cases respectively.  
The terms in the first row of the RHS of Eq.~\eqref{eq:chisq} correspond to the statistical contribution. The first two terms correspond to the algebraic difference while the last term corresponds to the fractional difference between the  two sets of events.
Note that $p^{\text{data}}$ and $p^{\text{fit}}$ refer to the set of oscillation parameters for the calculation of $N^{(3+0)}$ and $N^{(3+1)}$ respectively.  $p^{\text{data}}:\{\ta, \tb, \tc, \da, \sdm, \ldm \}$ and $p^{\text{fit}}:\{\ta, \tb, \tc, \td, \te, \tf, \da, \db, \dc, \sdm, \ldm, \lldm \}$. $\sigma_{p_{l}}$ is the uncertainty in the prior measurement of  $p_{l}$. 
In order to generate events in the $(3+0)$ case (referred  to as the \textit{data}), we fix the values of the six standard
oscillation parameters at their best-fit values. The values of the best-fit oscillation parameters and their uncertainties  are listed in Table~\ref{tab:parameters}.  In order to compute events in the ${(3+1)}$ case, we take fixed value of $\lldm = 1 \text{ eV}^{2}$ and vary the unknown parameters from the set $p^{\text{fit}}$\,
\footnote{
Note that the definition of $\chisq$ described in Eq.\ \ref{eq:chisq} is Poissonian in nature. In the limit of large events,  this reduces to the Gaussian form : 
\begin{align*}
 \Delta \chi^{2}(\bar{p}^{\text{fit}})  = \underset{(p^{\text{fit}}-\bar{p}^{\text{fit}}; \eta)}{{\text{min}}} 
\Bigg[\sum_{x}^{\text{mode}}\sum_{j}^{\text{channel}}\sum_{i}^{\text{bin}}
\frac{
\Big(
N_{ijxy}^{(3+1)}(p^{\text{fit}};\eta) - N_{ijxy}^{(3+0)}(p^{\text{data}})  
\Big)^{2}}{N_{ijxy}^{(3+0)}(p^{\text{data}})}
+ \text{prior} + \text{systematics}\Bigg].
\end{align*}}.
 \begin{table}[h]
\centering
\begin{adjustbox}{width=\textwidth}
\begin{tabular}{| c | c | c | c | c |}
\hline
&&&&\\
Usage & Systematics ($\eta_{k}$) & Value ($\sigma_{k}$) & Impacts & ND/FD correlated?  \\
&&&&\\
\hline
&&&&\\
\multirow{9}{2cm}{FD-only Analysis} & $\nu_{e}$-signal & 0.02 &  All events from $\nu_{e}$-signal at FD 
& \multirow{10}{*}{$-$} \\
& $\bar{\nu}_{e}$-signal & 0.02 &  All events from $\bar{\nu}_{e}$-signal at FD &\\
& $\nu_{\mu}$-signal & 0.05 &  All events from $\nu_{\mu}$-signal at FD &\\
& $\bar{\nu}_{\mu}$-signal & 0.05 &  All events from $\bar{\nu}_{\mu}$-signal at FD &\\
& $\nu_{\mu}$-background & 0.05 &  All events from $\nu_{\mu}$-background at FD &\\
& $\nu_{e}$-background & 0.05 &  All events from $\nu_{e}$-background at FD &\\
& $\bar{\nu}_{e}$-background & 0.05 &  All events from $\bar{\nu}_{e}$-background at FD &\\
& NC event & 0.1 &  All events from NC channel at FD &\\
&&&&\\
\hline
&&&&\\
\multirow{10}{2cm}{FD+ND Analysis} & FD Fiducial volume             & 0.01                    &  All events at FD   &  no\\
& ND Fiducial volume             & 0.01                    &  All events at ND   &  no\\
& Flux signal component    & 0.08  &  All events from signal component at FD \& ND   &  yes\\
& Flux background component    & 0.08  &  All events from background component at FD \& ND   &  yes\\
& Flux signal component n/f    & 0.004  &  All events from signal component at ND   &  no\\
& Flux background component n/f    & 0.02  &  All events from background component at ND   &  no\\
& CC cross-section (each flavour)    & 0.15  &  All CC events of each flavour at FD \& ND   &  yes\\
& NC cross-section    & 0.25  &  All NC events at FD \& ND   &  yes\\
& CC cross-section (each flavour)  n/f  & 0.02  &  All CC events of each flavour at ND   &  no\\
& NC cross-section  n/f  & 0.25  &  All NC events at ND   &  no\\
&&&&\\
\hline
\end{tabular}
\end{adjustbox}
\caption{\footnotesize{\label{tab:systematics}
 The table lists the systematics parameters $\eta_{k}$ and their values (uncertainties or $\sigma_{k}$) used in our analysis. 
 The first column indicates whether the corresponding systematics have been used for FD-only analysis or for (FD+ND)-analysis.
 The second column shows the list of systematic parameters while the third column shows the corresponding normalization uncertainties. 
 The last two columns indicates which events are impacted by the respective systematics and whether the systematic parameter is correlated between FD and ND. 
 The values for the FD-only systematics are chosen from \cite{DUNE:2021cuw}, while that for (FD+ND)-analysis are chosen from \cite{DUNE:2020fgq}.
}}
\end{table}

The two terms in the second row correspond to the prior and systematics respectively. 
The \textit{prior} term accounts for the penalty of the $l$ 
number of \textit{fit} parameters deviating away from the 
corresponding $p^{\text{data}}$. 
The degree of this deviation is controlled by  $\sigma_{pl}$ which is the uncertainty in 
the prior measurement of the best-fit values of $p^{\text{data}}$. 
The  \textit{systematics}-term accounts for the variation of the systematic/nuisance parameters.
$\eta$ is the set of values of $k$-systematics parameters $\{\eta_{1}$, $\eta_{2}$, $\dots$ $\eta_{k}\}$  while 
$\sigma_k$ is the uncertainty in the corresponding  systematics. 

This way of treating the nuisance parameters in the $\chisq$ calculation is known as the {\it{method of pulls}}~\cite{Huber:2002mx,Fogli:2002pt,GonzalezGarcia:2004wg,Gandhi:2007td}. 
For the analysis of events simulated at DUNE FD, we use the multiplicative systematical uncertainties included within the GLoBES configuration files~\cite{DUNE:2021cuw}. 
These uncertainties already take into account the rate constraints imposed by the ND~\cite{DUNE:2020fgq}. 
For the analysis of the combined data at FD+ND together we have explicitly made a GLoBES glb file for the ND together with that for the FD after following the prescription of the DUNE collaboration~\cite{DUNE:2020fgq}. 
Thus for the (FD+ND) analysis, we considered a larger number of systematical errors, partially correlated between FD and ND. In Table \ref{tab:systematics}, we list all the systematic parameters 
as used in our analysis.
Further, for the (FD+ND)-analysis we multiply the existing smearing matrices in the GLoBES TDR configuration files with an additional Gaussian smearing (with $20\%$ energy resolution) and use the modified smearing for the calculation~\cite{DUNE:2020fgq}, which is obtained by integrating (over the reconstructed energy $E'$) the Gaussian,
\begin{align}
R^{c}(E, E')= \dfrac{1}{\sigma(E) \sqrt{2 \pi}}\,\text{exp}\bigg[{-\dfrac{(E-E')^{2}}{2{(\sigma(E))}^{2}}}\bigg]\,, \nonumber
\end{align} 
with $\sigma(E) = 0.2E$, where $E$ is the true neutrino energy from simulation.
The final estimate of the $\chisq$ as a function of the set of desired parameters ($\bar{p}^{\text{fit}}$) for a given set of fixed parameters $p^{\text{data}}$ is obtained 
after minimizing the entire quantity within the square bracket in Eq.\ \eqref{eq:chisq} over the relevant set of rest of the \textit{fit} parameters $p^{\text{fit}}-\bar{p}^{\text{fit}}$, as well as over the systematics $\eta$.
This minimization is also referred to as marginalization over the set $\{p^{\text{fit}}-\bar{p}^{\text{fit}}; \eta\}$.  
Technically, this  procedure is the frequentist method of hypotheses testing~\cite{Fogli:2002pt, Qian:2012zn}.

\begin{figure}[t!]
\includegraphics[scale=0.65]{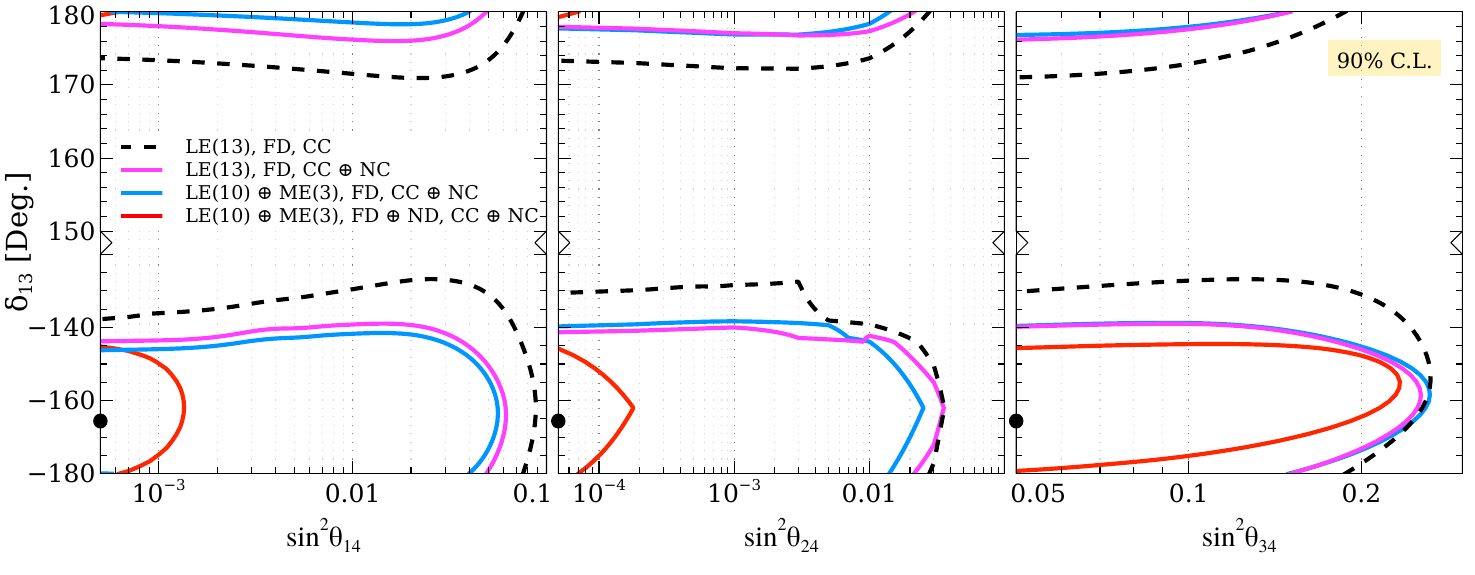}
\caption{\footnotesize{
The figure shows the $\chisq$ contours at $90\%$ C.L. in the parameter space of $\sin^{2}\theta_{i4}-\da$ ($i=1,2,3$ in the three panels respectively) for various analysis configurations. 
The analysis modes considered are: LE
beam at the FD using the CC interaction of neutrinos (black dashed); LE beam at the FD using the CC and NC interactions (magenta dotted);  
a combination LE beam and ME beam at the FD using CC and NC interactions (blue solid); a combination of LE beam and ME beam at the FD angumented with the corresponding to ND analysis by using both CC and NC interactions of neutrino (red solid). 
The numbers in parantheses beside each legend indicates the corresponding total runtime (in years) shared 
equally in $\nu$ and $\bar{\nu}$ modes.
Note that the $\da$ axis is broken (\textit{i.e.}, not shown) in between $-120^{\circ}$ and $150^{\circ}$ for ease of visibility of the contours.
The black dot indicates the fixed best-fit value of 
$\da$ considered in the simulated \textit{data}.
}}
\label{fig:chisq_dcp_thi4}
\end{figure}
\section{Results}
\label{sec:results}

We consider the following  configurations (beam tune, runtime,  channel, detector) for the sensitivity analysis in the present section.
\begin{enumerate}
\item[(a)]  LE ($13$), FD, CC
\item[(b)] LE ($13$), FD, CC $\oplus$ NC
\item[(c)] LE ($10$) $\oplus$ ME ($3$), FD, CC $\oplus$ NC
\item[(d)]  LE ($10$) $\oplus$ ME ($3$), FD $\oplus$ ND, CC $\oplus$ NC
\end{enumerate}
The numbers in parantheses beside each combination indicates the corresponding total runtime (in years) shared 
equally in $\nu$ and $\bar{\nu}$ modes.
The four configurations are depicted by  
(a)  black dotted curve, (b)  magenta solid curve, 
(c)  blue solid curve, and 
(d)  red solid curve  in Fig.\ \ref{fig:chisq_dcp_thi4}, Fig.\ \ref{fig:chisq_th23_thi4} and Fig.\ \ref{fig:chisq_dm41_thi4}.

The relevant oscillation channels are : $\nu_\mu \to \nu_e$, $\nu_\mu \to \nu_\mu$, $\nu_\mu \to \nu_\tau$ and $\nu_\mu \to \nu_s$ along with the corresponding anti-neutrino channels. In Fig.\ \ref{fig:chisq_dcp_thi4} and  Fig.\ \ref{fig:chisq_th23_thi4}, we show the $\chisq$ contours at $90\%$ confidence level (C.L.) (which corresponds to $\chisq = 4.6$ for the case of two-parameters~\cite{ParticleDataGroup:2022pth}) for  fixed value of $\lldm = 1 \text{ eV}^{2}$ in 
the parameter space of $(\sin^{2}\theta_{i4}-\da)$ 
and $(\sin^{2}\theta_{i4}-\tc)$ respectively, where   $i=1,2,3$.  
In our sensitivity analysis, we marginalise over  the set of parameters $\{|\ldm|, \tc, \da, \td, \te, \tf, \db, \dc\}$ while leaving out those parameters that are depicted on the axes.
Thus, for instance, in the middle panel of Fig.\ \ref{fig:chisq_dcp_thi4}, the full set of marginalised parameters will be $\{|\ldm|,  \tc, \td, \tf, \db, \dc \}$. The parameter space lying outside the contours are 
excluded at $90\%$ C.L.
In Fig.\ \ref{fig:chisq_dm41_thi4}, we depict $\chisq$ contours at $90\%$ C.L. in the 
parameter space of $(\sin^{2}\theta_{i4}-\lldm)$ where $i=1,2$~\footnote{We have checked that the parameter space $(\sin^{2}\tf-\lldm)$ cannot be excluded efficiently thereby showing no visible contours in the relevant parameter range due to low sensitivity to $\tf$. 
Hence, we did not include the corresponding figure here. 
Note that, unlike previous figures,  we varied $\lldm$ in Fig.\ \ref{fig:chisq_dm41_thi4} and the total set of marginalised parameters includes $\{|\ldm|, \db, \dc, \tc, \da, \tf\}$.
In generating Fig.\ \ref{fig:chisq_dm41_thi4}, we have taken a stepsize of 0.001 (0.0001)  for $\sin^{2}\td$ ($\sin^{2}\te$). Along the $\lldm$ axis, the stepsize is 0.001 eV$^{2}$ in the range $[0.001, 0.01]$ eV$^{2}$;  0.01 eV$^{2}$ in the range $[0.01, 0.1]$ eV$^{2}$ and so on.
}. 
 Below, we summarize the key observations from 
Fig.\ \ref{fig:chisq_dcp_thi4},  Fig.\ \ref{fig:chisq_th23_thi4} and Fig\ \ref{fig:chisq_dm41_thi4}.

\begin{itemize}
\item
In Fig.\ \ref{fig:chisq_dcp_thi4} and Fig.\ \ref{fig:chisq_th23_thi4}, we note that  
$\sin^{2}\td$ and $\sin^{2}\te$ can be 
 constrained much better than $\sin^{2}\tf$ (by almost an order of magnitude) 
and the constraints for $\sin^{2}\te$ are tighter than those for $\sin^{2}\td$ and $\sin^{2}\tf$ (This is consistent with, for instance Fig. 3 of \cite{Singha:2022btw}).

\item
Comparing the various analysis configurations in Fig.\ \ref{fig:chisq_dcp_thi4} and Fig.\ \ref{fig:chisq_th23_thi4}, we note that rather than using LE beam only (a), sharing the runtime with ME beam  (c)
offers a slight improvement for $\sin^{2}\te$, but very little for $\sin^{2}\td$. 
Using the NC channel along with the usual CC channel (b)  
also tightens the constraints on $\sin^{2}\theta_{i4}$ ($i=1,2,3$) slightly. 
However, using the ND data along with the FD  (d)
leads to significant improvement over the other configurations and leads to stringent constraints on 
$\sin^{2}\td$ and $\sin^{2}\te$.
This improvement is almost two (one) orders of magnitude for $\sin^{2}\te$ ($\sin^{2}\td$). 
In Table~\ref{tab:chisq_constraints}, we summarize these observations in a quantitative manner.
The table shows the minimum (\ie, most conservative) 
estimate of $\sin^{2}\theta_{i4}$ ($i=1,2,3$) that 
can be ruled out by the given configuration at $90\%$ C.L. 
The values in Table~\ref{tab:chisq_constraints} are obtained from Fig.\ \ref{fig:chisq_dcp_thi4}. It should be noted that the results in Fig.\ \ref{fig:chisq_th23_thi4} are consistent with Table~\ref{tab:chisq_constraints}.
%
\item All the (red) contours in Fig.\ \ref{fig:chisq_dcp_thi4}, Fig.\ \ref{fig:chisq_th23_thi4} and Fig.\ \ref{fig:chisq_dm41_thi4}  show that the sterile parameters are strongly constrained when  we consider option (d).
Since the oscillation effect due to $\lldm \sim \mathcal{O}(1 \text{ eV}^{2})$ is most pronounced at baselines $\lesssim 1 $ km, the ND can probe the sterile parameters very efficiently. 
In Fig.\ \ref{fig:chisq_dm41_thi4}, 
we find that varying $\lldm$, the most stringent exclusion region on the mixing angles are
\begin{figure}[t!]
\includegraphics[scale=0.65]{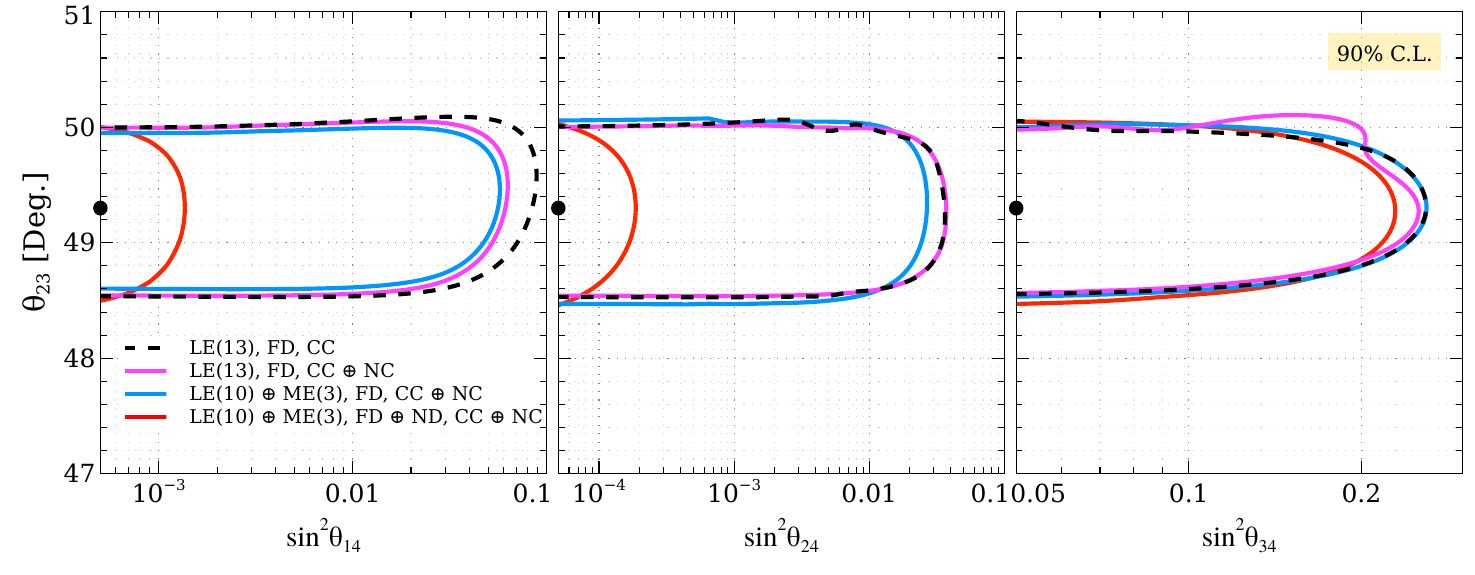}
\caption{\footnotesize{
Similar to Fig.\ \ref{fig:chisq_dcp_thi4} but showing the $\chisq$ contours in the parameter space of 
$\sin^{2}\theta_{i4}-\tc$ ($i=1,2,3$ in the three panels respectively).
}}
\label{fig:chisq_th23_thi4}
\end{figure}
\begin{align}
\sin^{2}\td &\gtrsim 2 \times 10^{-3}\, (90 \%\, \text{C.L.}) 
\text{ at } \lldm \simeq 11 \text{ eV}^{2}\,, \nonumber \\
{\sin^{2}\te} &\gtrsim 3 \times 10^{-4}\, (90 \%\, \text{C.L.})
\text{ at } \lldm \simeq 7 \text{ eV}^{2}.\nonumber
\end{align}
%
\end{itemize}
We can get a qualitative sense of different configurations as follows.
Deconstructing the $\chisq$ quantity, we obtain,
\begin{align}
\chisq  &\simeq 
\chisq_{\mu e}
+ \chisq_{\mu \mu} \nonumber \\
&\sim \sum_{\text{bin}}\bigg[\frac{(N_{\mu e}^{(3+1)}-N_{\mu e}^{(3+0)})^{2}}{N_{\mu e}^{(3+0)}}
+ \frac{(N_{\mu \mu}^{(3+1)}-N_{\mu \mu}^{(3+0)})^{2}}{N_{\mu \mu}^{(3+0)}}\bigg] \nonumber \\
&\simeq \sum_{\text{bin}}\bigg[
\frac{|\Delta P_{\mu e}|^{2}}{P^{(3+0)}_{\mu e}}\sigma_{\nu_{e}}\Phi_{\nu_{\mu}}
+ \frac{|\Delta P_{\mu \mu}|^{2}}{P^{(3+0)}_{\mu \mu}}\sigma_{\nu_{\mu}}\Phi_{\nu_{\mu}}\bigg],
\end{align}
where $N^{(3+1)}_{\alpha\beta}$ and $N^{(3+0)}_{\alpha\beta}$ are the event spectra 
corresponding to $\nu_{\alpha} \to \nu_{\beta}$ 
channel 
for the $(3+1)$ and the $(3+0)$ scenario respectively. 
$\Delta P_{\alpha \beta} = P_{\alpha \beta}^{(3+0)} 
- P_{\alpha \beta}^{(3+1)}$, as discussed in Section.\ \ref{sec:chisq}. 
$\Phi_{\nu_\alpha}$ and $\sigma_{\nu_\beta}$   ($\alpha, \beta = e, \mu, \tau$) are the flux and cross-sections of neutrino flavour $\nu_{\alpha}$ 
and $\nu_{\beta}$ respectively. Note that here we schematically~\footnote{The actual $\chisq$ will also get contribution from the systematics and prior term, as explained in Eq.\ \eqref{eq:chisq} and the related discussions. Here 
we analyse only the dominant statistical contribution. We also ignore the detector response factors.} write the contribution to $\chisq$ from 
the two oscillation channels $\mue$ and $\mumu$. 
Since the $\bar{\nu}$ contribution to $\chisq$ is 
expected to be much less, we analyse only the 
$\nu$ running mode. We can write the $\chisq$ for the four analysis configurations relevant for the present article.
\begin{align}
\chisq\big[\text{LE ($13$), FD, CC 
}\big] &\sim 
\sum_{\text{bin}}
\bigg[
\bigg(
\frac{|\Delta P_{\mu e}|^{2}}{P_{\mu e}^{(3+0)}}\sigma_{\nu_{e}}^{CC}
+ \frac{|\Delta P_{\mu \mu}|^{2}}{P_{\mu \mu}^{(3+0)}}\sigma_{\nu_{\mu}}^{CC}
\bigg)
13\Phi_{\nu_{\mu}}^{LE}
\bigg], 
\label{eq:chisq_le} \\[3mm]
\chisq\big[\text{LE ($13$), FD, CC $\oplus$ NC
}\big] &\sim 
\sum_{\text{bin}}
\bigg[
\bigg(
\frac{|\Delta P_{\mu e}|^{2}}{P_{\mu e}^{(3+0)}}\sigma_{\nu_{e}}^{CC}
+ \frac{|\Delta P_{\mu \mu}|^{2}}{P_{\mu \mu}^{(3+0)}}\sigma_{\nu_{\mu}}^{CC}
+ \frac{|\Delta P_{NC}|^{2}}{P_{NC}^{(3+0)}}\sigma^{NC}
\bigg)
13\Phi_{\nu_{\mu}}^{LE}
\bigg], 
\label{eq:chisq_le_nc} \\[3mm]
\chisq\big[\text{LE ($10$) $\oplus$ ME ($3$), FD, CC $\oplus$ NC
}\big] &\sim 
\sum_{\text{bin}}
\bigg[
\bigg(
\frac{|\Delta P_{\mu e}|^{2}}{P_{\mu e}^{(3+0)}}\sigma_{\nu_{e}}^{CC}
+ \frac{|\Delta P_{\mu \mu}|^{2}}{{P_{\mu \mu}^{(3+0)}}}\sigma_{\nu_{\mu}}^{CC}+\frac{|\Delta P_{NC}|^{2}}{P_{NC}^{(3+0)}}\sigma^{NC}
\bigg) \nonumber  \\ &
\bigg(
10\Phi_{\nu_{\mu}}^{LE} 
+  3\Phi_{\nu_{\mu}}^{ME}
\bigg)
\bigg], 
\label{eq:chisq_le_me_nc} 
\end{align}
\begin{align}
\chisq\big[\text{LE ($10$) $\oplus$ ME ($3$), FD $\oplus$ ND, CC $\oplus$ NC
}\big] &\sim 
\sum_{\text{bin}}
\bigg[
\bigg\{
\bigg(
\frac{|\Delta P_{\mu e}|^{2}}{P_{\mu e}^{(3+0)}}\sigma_{\nu_{e}}^{CC}
+ \frac{|\Delta P_{\mu \mu}|^{2}}{P_{\mu \mu}^{(3+0)}}\sigma_{\nu_{\mu}}^{CC}
+ \frac{|\Delta P_{NC}|^{2}}{P_{NC}^{(3+0)}}\sigma^{NC}
\bigg)
\nonumber \\
&+
\frac{M_{\text{ND}}}{M_{\text{FD}}}
\frac{L_{\text{FD}}^{2}}{L_{\text{ND}}^{2}}
\bigg(
\frac{|\Delta P_{\mu e}^{\text{ND}}|^{2}}{P_{\mu e}^{(3+0), \text{ND}}}\sigma_{\nu_{e}}^{CC}
+ \frac{|\Delta P_{\mu \mu}^{\text{ND}}|^{2}}{P_{\mu \mu}^{(3+0), \text{ND}}}\sigma_{\nu_{\mu}}^{CC} \nonumber
\\&+\frac{|\Delta P_{NC}^\text{ND}|^{2}}{P_{NC}^{(3+0),\text{ND}}}\sigma^{NC}
\bigg)
\bigg\} 
\bigg(
10\Phi_{\nu_{\mu}}^{LE} 
+ 3\Phi_{\nu_{\mu}}^{ME}
\bigg)
\bigg].
\label{eq:chisq_le_me_nc_nd}
\end{align}

Unless otherwise mentioned, all probabilities are calculated at the FD location of DUNE with $L = L_{\text{FD}} = 1300$ km. 
Only the probabilities with explicit label ``ND" (in Eq.\ \eqref{eq:chisq_le_me_nc_nd}) are 
calculated at the ND location of DUNE with $L = L_{\text{ND}} = 0.57$ km.
Note that, for fair comparison, we have multiplied the runtimes (in years) with the 
respective flux terms. 
Furthermore, in Eq.\ \eqref{eq:chisq_le_me_nc_nd}, the 
factor $L_{\text{FD}}^{2}/L_{\text{ND}}^{2}$ is multiplied with the ND-term in 
order to convey that the flux at FD location 
is approximately reduced by a factor $L_{\text{FD}}^{2}/L_{\text{ND}}^{2}$ compared to that at ND location. 
We have also considered the difference of the 
fiducial masses of the ND ($M_{\text{ND}} \simeq 0.067$ kt) and FD ($M_{\text{FD}} \simeq 40$ kt) and multiplied the ND-term by their ratio.

Among the fractional probability difference terms ($|\Delta P_{\alpha \beta}|/P_{\alpha \beta}^{(3+0)}$), we have already seen that 
$|\Delta P_{\mu e}|/P_{\mu e}^{(3+0)}$ behaves in a 
qualitatively similar manner for $\td$ and $\te$ 
and manifests substantial deviation that is roughly uniform throughout the energy range
(see Fig.\ \ref{fig:delta_p} and the related discussions). 
On the other hand, though $|\Delta P_{\mu \mu}|/P_{\mu \mu}^{(3+0)}$ can give some contribution 
in case of $\te$, it is practically insensitive to 
$\td$ (Fig.\ \ref{fig:delta_p}). 
For a given analysis 
configuration, $\sin^{2}\te$ can thus be constrained 
better than $\sin^{2}\td$ since 
 the $\mumu$ channel (in addition to the 
 $\mue$ channel) also contributes 
significantly in that case. 

One can note that the  LE beam ($\Phi_{\nu_{\mu}}^{LE}$) peaks around energy range $2-3$ GeV and falls rapidly beyond roughly $5$ GeV. 
On the other hand, the ME beam is significant till around $\gtrsim 5$ GeV while being of slightly less (roughly half) magnitude than LE beam around $2-3$ GeV (Fig.\ \ref{fig:flux_tdr}).
Hence, in the low-energy bins (upto $3$ GeV), the major contribution to
$\chisq$ in Eq.\ \eqref{eq:chisq_le_me_nc} 
comes from the term $(|\Delta P_{\mu e}|/P_{\mu e}^{(3+0)})\Phi_{\nu_{\mu}}^{LE}$. 
 Fig.\ \ref{fig:delta_p} also shows that  $|\Delta P_{\mu \mu}|/P_{\mu \mu}^{(3+0)}$ 
is practically insensitive to $\td$ for all energies and is also 
close to zero in case of $\te$ for $E \simeq 2-4$ GeV. This channel 
 does not contribute significantly to $\chisq$ for either $\sin^{2}\td$ or $\sin^{2}\te$ in the low energy range.
For higher energy bins ($\gtrsim 5$ GeV), the main 
contribution to $\chisq$ in Eq.\ \eqref{eq:chisq_le_me_nc} comes from 
$(|\Delta P_{\mu e}|^{2}/P_{\mu e}^{(3+0)}+|\Delta P_{\mu \mu}|^{2}/P_{\mu \mu}^{(3+0)})\Phi_{\nu_{\mu}}^{ME}$ for 
$\sin^{2}\te$ and only from $(|\Delta P_{\mu e}|^{2}/P_{\mu e}^{(3+0)})\Phi_{\nu_{\mu}}^{ME}$ for 
$\sin^{2}\td$.
This explains why the contours in Fig.\ \ref{fig:chisq_dcp_thi4}, Fig.\ \ref{fig:chisq_th23_thi4} and Fig.\ \ref{fig:chisq_dm41_thi4} shrink slightly 
for $\sin^{2}\te$ (but not for $\sin^{2}\td$) when the runtime is distributed among the LE and ME. 
Including the NC channel increases the $\chisq$ 
slightly and this can be clearly understood 
by the presence of the additional term $(|\Delta P_{NC}|^{2}/P_{NC}^{(3+0)})\sigma_{NC}$ in Eq.\ \eqref{eq:chisq_le_nc} when compared to Eq.\ \eqref{eq:chisq_le}.

Analysing Eq.\ \eqref{eq:chisq_le_me_nc_nd}, we observe 
that the dominant contribution to the corresponding 
$\chisq$ comes from the two terms proportional to 
$|\Delta P_{\mu e}|^{2}/P_{\mu e}^{(3+0)}$ 
and 
$(M_{\text{ND}}/M_{\text{FD}})(L_{\text{FD}}/L_{\text{ND}})^{2}\big(|\Delta P_{\mu e}^{\text{ND}}|^{2}/P_{\mu e}^{(3+0), \text{ND}}\big)$. 

If $L_{\text{FD}}$ and $L_{\text{ND}}$ are the baselines corresponding to the FD and ND location of DUNE, then the probabilities at FD and ND are mainly driven by the oscillation terms $\sin^{2}(\Delta m^{2}_{31}L_{\text{FD}}/4E)$ and $\sin^{2}(\Delta m^{2}_{41}L_{\text{ND}}/4E)$ respectively.
Thus, for an order-of-magnitude estimation of the comparative impacts of FD and ND, the calculation boils down to the comparison of 
$\sin^{2}(\Delta m^{2}_{31}L_{\text{FD}}/4E)$ 
and $(M_{\text{ND}}/M_{\text{FD}})(L_{\text{FD}}/L_{\text{ND}})^{2}\sin^{2}(\Delta m^{2}_{41}L_{\text{ND}}/4E)$ 
respectively. 
The calculation of $\chisq$ is proportional to 
$\Phi_{\nu_{\mu}}^{\text{LE}}$ (see Eq.\ \eqref{eq:chisq_le_me_nc_nd}), which peaks around $2-3$ GeV, - making it the dominant energy range which drives the calculation of $\chisq$.
Assuming $L_{\text{FD}} \simeq 1.3 \times 10^{3}$ km,
in that energy range, 
\begin{equation}
\label{eq:dm31_fd}
\sin^{2}(\Delta m^{2}_{31}L_{\text{FD}}/4E) \sim 1, 
\end{equation}
since the argument of \textit{sine} is close to $\pi/2$. 

Assuming $L_{\text{ND}} \simeq 0.57$ km, 
$M_{\text{FD}} \simeq 40$ kt, 
$M_{\text{ND}} \simeq 0.067$ kt,
\begin{align}
\label{eq:dm41_nd}
\frac{M_{\text{ND}}}{M_{\text{FD}}}
\frac{L_{\text{FD}}^{2}}{L_{\text{ND}}^{2}}\sin^{2}\bigg(\frac{\Delta m^{2}_{41}L_{\text{ND}}}{4E}\bigg) 
&\sim 8.7 \times 10^{3}
\sin^{2}\bigg(
1.27\,\frac{\Delta m^{2}_{41} [\text{eV}^{2}] L_{\text{ND}}[\text{km}]}{E [\text{GeV}]} \bigg) \nonumber \\
&\sim 8.7 \times 10^{3} \sin^{2}\bigg(0.72\,\frac{\Delta m^{2}_{41} [\text{eV}^{2}]}{E[\text{GeV}]}\bigg)
\end{align}
\begin{figure}[t!]
\centering
\includegraphics[scale = 0.7]{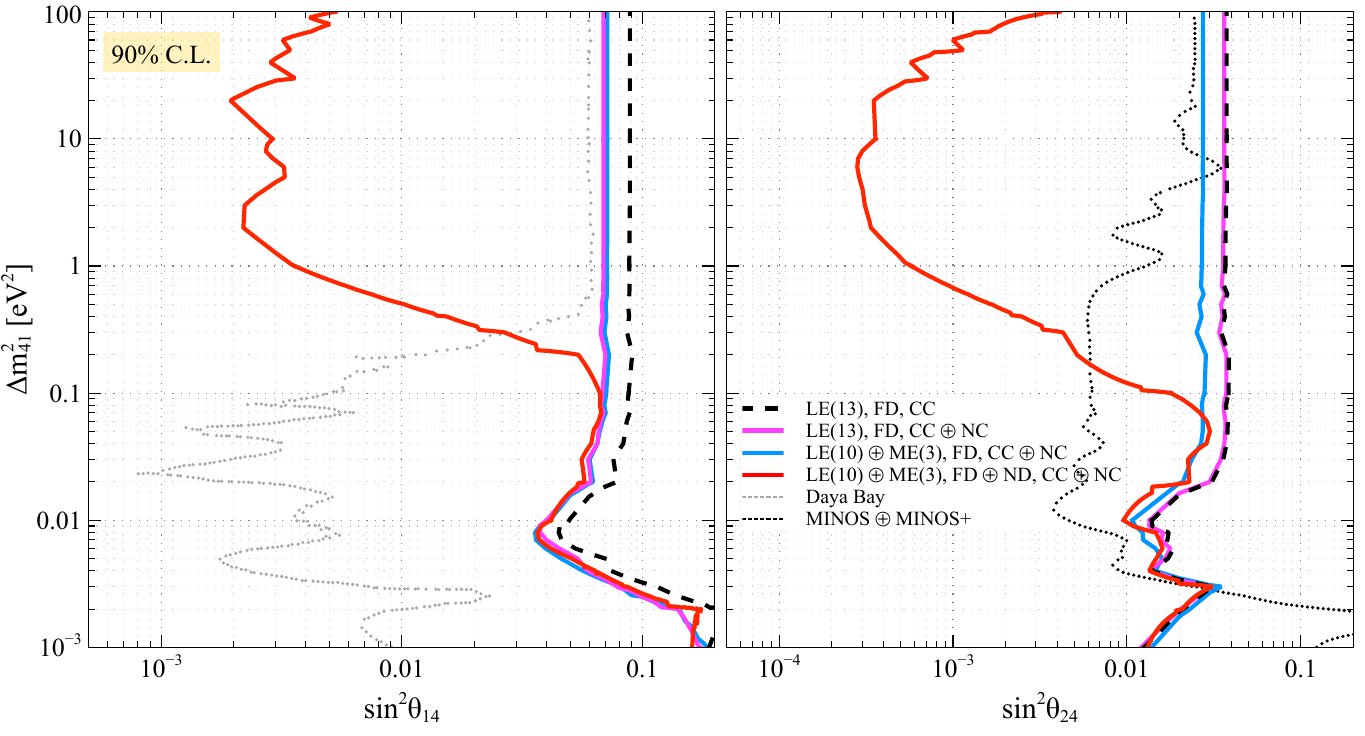}
\caption{\footnotesize{
The figure shows the $\chisq$ contours at $90\%$ C.L. in the 
parameter space of $\sin^{2}\theta_{i4}-\lldm$  ($i=1,2$ in the two panels respectively) for 
various analysis configurations. 
The configurations considered are: LE
beam at the FD using the CC interaction of neutrinos (black dashed); LE beam at the FD using the CC and NC interactions (magenta dotted);  
a combination LE beam and ME beam at the FD using CC and NC interactions (blue solid); a combination of LE beam and ME beam at the FD angumented with the corresponding to ND analysis by using both CC and NC interactions of neutrino (red solid) The numbers in parantheses beside each legend indicates the corresponding total runtime (in years) shared 
equally in $\nu$ and $\bar{\nu}$ modes. {For the comparison Daya Bay's $90\%$ C.L.\ \cite{DayaBay:2016lkk} in a plane of $\sin^{2}\theta_{14}-\lldm$  (dotted gray in the left panel of figure) and MINOS $\&$ MINOS+ at $90\%$ C.L.\ \cite{MINOS:2017cae} in a plane of $\sin^{2}\theta_{24}-\lldm$  (dotted black in the right panel of figure) .}}}
\label{fig:chisq_dm41_thi4}
\end{figure}
The contribution by ND is maximum when the argument 
of $sine$ is roughly $\pi/2$ in Eq.\ \eqref{eq:dm41_nd}. 
This implies $\Delta m^{2}_{41} [\text{eV}^{2}] \sim (2.3\,E \,[\text{GeV}])$. 
Thus, for the dominat energy range of $2-3$ GeV, ND can constrain the sterile mixing angles most efficiently around $\lldm \simeq 4-7$ $\text{eV}^{2}$. 
These numbers are close to our observations in Fig.\ \ref{fig:chisq_dm41_thi4}.
Slight deviations may arise due to the fact that we have ignored higher order terms and the contribution of $\mumu$ disappearance channel  in our simplified analyses. 
\begin{table}
    \centering
   \begin{tabular}{|l|c|c|c|c|}
    \hline
       \multirow{2}{*}{Analysis configurations}
       & \multicolumn{3}{|c|}{Active-Sterile mixing angles} \\
        \cline{2-4}
        & $\sin^{2}\theta_{14}$ & $\sin^{2}\theta_{24}$ & 
         $\sin^{2}\theta_{34}$ \\ \hline
        LE ($13$), FD, CC
 & $8.77 \times 10^{-2}$ & $3.55 \times 10^{-2}$ & $2.65 \times 10^{-1}$ \\ \hline
         LE ($13$), FD, CC $\oplus$ NC
& $6.20 \times 10^{-2}$ & $3.52 \times 10^{-2}$ & $2.63 \times 10^{-1}$ \\ \hline
          LE ($10$) $\oplus$ ME ($3$), FD, CC $\oplus$ NC
 & $5.62 \times 10^{-2}$  & $2.49 \times 10^{-2}$ & $2.53 \times 10^{-1}$ \\ \hline
          LE ($10$) $\oplus$ ME ($3$), FD $\oplus$ ND, CC $\oplus$ NC
 & $1.4 \times 10^{-3}$ & $1.8 \times 10^{-4}$ & $2.34 \times 10^{-1}$ \\ \hline
    \end{tabular}
    \caption{\footnotesize{The table shows the minimum value of $\sin^{2}\theta_{i4}$ ($i=1,2,3$) that can be ruled out at $90\%$ C.L. 
    irrespective of the value of $\da$ from Fig.\ \ref{fig:chisq_dcp_thi4}.}}
    \label{tab:chisq_constraints}
\end{table}
\section{Conclusion}
\label{sec:conclusion}
DUNE is a very promising next generation long baseline neutrino experiment as it has a remarkable capability to probe the standard oscillation parameters with unprecedented precision and to shed light on the unknowns in the standard three flavor paradigm ($CP$ violating phase, $\delta_{13}$, octant of $\theta_{23}$ and neutrino mass hierarchy)~\cite{DUNE:2020ypp}.  DUNE also has the ability to constrain the sub-dominant new physics scenarios such as non-standard interaction (NSI), sterile neutrino, decoherence etc~\cite{DUNE:2020fgq,Farzan:2017xzy,Proceedings:2019qno}. 
In the present work, we focus on analysing the capability of DUNE to study the correlations among the active-sterile  parameters and to constrain the sterile parameters. 

In order to study the role played by eV-scale sterile neutrinos, we compute the probability difference between the $(3+0)$ and  $(3+1)$ case 
for the three relevant channels: $\nu_{\mu} \to \nu_{e}$ channel, $\nu_{\mu} \to \nu_{\mu}$ channel and the NC channel.
We numerically calculate $|\Delta P_{\alpha \beta}|/P_{\alpha \beta}^{(3+0)}$ as a function of energy and study the  impact of three active-sterile mixing angles ($\td, \te, \tf$) on this quantity using  simplified analytic expressions. 
We find that $|\Delta P_{\mu e}|/P_{\mu e}^{(3+0)}$ is impacted the same way as we increase $\td$ or $\te$ with almost uniform increase in magnitude across all energies under consideration. 
Both $|\Delta P_{\mu\mu}|/P_{\mu \mu}^{(3+0)}$ and $|\Delta P_{\text{NC}}|/P_{NC}^{(3+0)}$ are practically independent of $\td$. 
But $\te$ gives rise to a prominent kink in $|\Delta P_{\mu\mu}|/P_{\mu \mu}^{(3+0)}$ around $2.5$ GeV and increases the magnitude of $|P_{\text{NC}}|/P_{NC}^{(3+0)}$ more uniformly across the energies. 

Then we discuss the two beam options at DUNE, namely the LE tuned flux and the $\nu_{\tau}$-optimized ME tuned flux and generate the corresponding event spectra for $(3+0)$ and $(3+1)$ cases at DUNE FD. 
As expected using the LE flux, we obtain higher statistics around the oscillation maximum ($2-3$ GeV) with events falling rapidly at higher energies. 
The ME flux leads to larger events at higher energies ($\gtrsim 5$ GeV) (see  Fig.\ \ref{fig:event_LE} and  Fig.\ \ref{fig:event_ME}).

We perform comparative  statistical analysis of the capabilities of DUNE (with a total $13$ years of runtime, $6.5$ years each in $\nu$ and $\bar{\nu}$-mode) to probe the sterile neutrino parameter space using four analysis configurations~\cite{DUNE:2020ypp,DUNE:2021cuw,DUNE:2020fgq}: 
\begin{enumerate}
\item[(a)]  LE ($13$), FD, CC.
\item[(b)] LE ($13$), FD, CC $\oplus$ NC
\item[(c)] LE ($10$) $\oplus$ ME ($3$), FD, CC $\oplus$ NC
\item[(d)]  LE ($10$) $\oplus$ ME ($3$), FD $\oplus$ ND, CC $\oplus$ NC
\end{enumerate}
We probe the parameter space, $\sin^{2}\theta_{i4}-\da, \sin^{2}\theta_{i4}-\tc$ and $\sin^{2}\theta_{i4}-\lldm$ $(i=1,2,3)$ using the 
four configurations listed above (see  Fig.\ \ref{fig:chisq_dcp_thi4}, Fig.\ \ref{fig:chisq_th23_thi4} and Fig.\ \ref{fig:chisq_dm41_thi4}). 
We find that in general $\sin^{2}\td$ and $\sin^{2}\te$ can be constrained much better (roughly by an order of magnitude) than $\sin^{2}\tf$, while 
$\sin^{2}\te$ can be constrained 
more strongly than $\sin^{2}\td$. 
Compared to the LE case, the configuration (LE $\oplus$ ME) improves the constraints only for $\sin^{2}\te$, whereas (LE, CC $\oplus$ NC) improves the constraints slightly for both $\sin^{2}\td$ and $\sin^{2}\te$. 
But the configuration (LE $\oplus$ ME, FD $\oplus$ ND, CC $\oplus$ NC) offers the most stringent constraints, $\sin^{2}\td\,(\sin^{2}\te) \lesssim 1.4 \times 10^{-3}\,(1.8 \times 10^{-4})$, which are one (two) orders of magnitude tighter compared to that estimated by the other configurations. 

Finally, our main result is contained in Fig.\ \ref{fig:chisq_dm41_thi4} which depicts the exclusion plot in the plane of $\sin^{2}\theta_{i4}-\lldm$ $(i=1,2)$ for the four considered configurations.  Using configuration (d) in the parameter space of $\sin^{2}\theta_{i4}-\lldm$ $(i=1,2)$ we estimate a greater tightening of the constraints for large $\lldm \gtrsim 0.01 \text{ eV}^{2}$ and estimate the most stringent constraint to be around $\lldm \simeq 3-10 \text{ eV}^{2}$. 
For comparison, the constraints from Daya Bay~\cite{DayaBay:2016lkk} and  MINOS and MINOS+~\cite{MINOS:2017cae} are also depicted in Fig.\ \ref{fig:chisq_dm41_thi4}. 
 Before we close,  we would like to make a few pertinent remarks about our analysis procedure. 
Note that in the high mass-splitting region, the reach of DUNE will be systematically limited. Our analysis for the ND has been carried out using the standard GLoBES procedure with systematics and normalization uncertainties from the TDR glb file.  Incorporating full systematic details (including shape-related uncertainties) for the ND in GLoBES will require  major modifications and is beyond the scope of the present study.  
\cleardoublepage

\appendix
\renewcommand{\theequation}{\thesection.\arabic{equation}}
\setcounter{equation}{0}
\renewcommand{\thesection}{\Alph{section}}
\section{Probability expression in $(3+0)$ and $(3+1)$ case in vacuum}
\label{appendix_a}
We provide the analytic expressions for the relevant probability channels in the $(3+0)$ and $(3+1)$. We adopt the parameterization as given in~\cite{Gandhi:2015xza}. Under the approximations given in Section\ \ref{sec:prob}, we have
\begin{align}
\label{p_mue_str}
P_{{\mu} e}^{(3+1)} (\te, \td)  &\simeq 
\dfrac{1}{2} \sin^{2} 2\theta_{13}  \cos^{2} \theta_{14} \cos^{2} \theta_{24}
\sin^2\dfrac{\Delta m^2_{31}L}{4E} \nonumber \\
&+ \frac{1}{\sqrt{2}}
\sin2\td \sin\te \cos\td \cos\te \sin 2\tb 
\sin \Big(\frac{\Delta m^2_{31}L}{4E} + \da \Big) 
\sin \frac{\Delta m^2_{31}L}{4E} \nonumber \\
&+\frac{1}{2}\sin^{2}2\td \sin^{2}\te 
+ \mathcal{O}(\lambda^{5})\,,
\\[3mm]
\label{p_mumu_str}
P^{(3+1)}_{{\mu}\mu}(\te, \td) &\simeq 1 - 2 \cos^{2}\theta_{14} \sin^{2} \theta_{24} -\Big[ \cos^{2} \theta_{13} + \cos^{2} \theta_{13} \sin^{2} \theta_{13} - 2 \cos^2\theta_{13} \sin^2\theta_{24} \Big]\sin^2\dfrac{\Delta m^2_{31}L}{4E} \nonumber \\ &+ \cos^2\theta_{13} \sin^2\theta_{24}\Big[\sin^2\theta_{13}  -\sin^{2}\theta_{24} -  2\sin^{2}\theta_{14}\cos^{2}\theta_{24} \Big] \sin^2\dfrac{\Delta m^2_{31}L}{4E}  \nonumber\\ &+ \mathcal{O}(\lambda^{5}) \,,
\\[3mm]
\label{p_nc_str}
P_{{NC}}^{(3+1)} (\theta_{14}, \theta_{24}, \theta_{34}) &\simeq 
\frac{1}{2} \cos^{4}\theta_{14} \cos^{2}\theta_{34} \sin^{2} 2\theta_{24} 
+ \cos\theta_{13}\cos^{2}\theta_{24}
\Big[\cos^3\theta_{13} \sin^2\theta_{34} 
\nonumber \\
&-\cos\theta_{13} \cos^2\theta_{34} \sin^2\theta_{24} 
+ \sqrt{2} \sin\theta_{13} \sin2\theta_{34}\sin\theta_{14} \cos\theta_{24} \Big] \sin^2\dfrac{\Delta m^2_{31}L}{4E} \nonumber \\
&+ \mathcal{O}(\lambda^{4}).
\end{align}
In the $(3+0)$ case, the simplified probability expressions (see Section\ \ref{sec:prob}) are
\begin{align}
\label{p_mue_std}
 P^{(3+0)}_{{\mu}e} &\simeq \frac{1}{2} \sin^{2} 2\theta_{13} \sin^2\dfrac{\Delta m^2_{31}L}{4E}\,,
 \\[3mm]
 \label{p_mumu_std}
 P^{(3+0)}_{{\mu}\mu}  &\simeq 1 - \Big[ \cos^{2} \theta_{13} + \cos^{2} \theta_{13} \sin^{2} \theta_{13} \Big] \sin^2\dfrac{\Delta m^2_{31}L}{4E}\,,\\[3mm]
 \label{p_nc_std}
P^{(3+0)}_{{NC}} &\simeq P_{\mu e} + P_{\mu \mu} + P_{\mu \tau} = 1\,.
\end{align}
\cleardoublepage

\section*{Acknowledgments}
We would like to thank Chris Marshall for useful comments. SP thanks Raj Gandhi for helpful discussion. SP acknowledges JNU for support in the form of fellowship. The  numerical analysis has been performed using the HPC cluster at SPS, JNU funded by DST-FIST. 
MM acknowledges support from the grant NRF-2022R1A2C1009686. 
MM also acknowledges the help of Kim Siyeon and Chang Hyon Ha for the usage of cluster at 
High Energy Physics Center in Chung-Ang University. This research (SP and PM) was supported in part by the International Centre for Theoretical Sciences (ICTS) for participating in the program - Understanding the Universe Through Neutrinos (code: ICTS/Neus2024/04).  This work reflects the views of the authors and not those of the DUNE collaboration.
\bibliographystyle{JHEP}
\bibliography{reference}

\providecommand{\href}[2]{#2}\begingroup\raggedright\begin{thebibliography}{100}

\bibitem{nobel2015}
T.~Kajita and A.B.~McDonald, ``For the discovery of neutrino oscillations,
  which shows that neutrinos have mass.''
  \url{https://www.nobelprize.org/prizes/physics/2015/summary/}, 2015.

\bibitem{10.5281/zenodo.4726908}
{P. F. De Salas}, {D. V. Forero}, {S. Gariazzo}, {P. Mart{\'\i}nez-Mirav{\'e}},
  {O. Mena}, {C. A. Ternes} et~al., ``{Chi2 profiles from Valencia neutrino
  global fit}.'' \url{http://globalfit.astroparticles.es/}, 2021.
\newblock 10.5281/zenodo.4726908.

\bibitem{deSalas:2020pgw}
P.F.~de~Salas, D.V.~Forero, S.~Gariazzo, P.~Mart{\'\i}nez-Mirav{\'e}, O.~Mena,
  C.A.~Ternes et~al., \emph{{2020 global reassessment of the neutrino
  oscillation picture}},
  \href{https://doi.org/10.1007/JHEP02(2021)071}{\emph{JHEP} {\bfseries 02}
  (2021) 071} [\href{https://arxiv.org/abs/2006.11237}{{\ttfamily
  2006.11237}}].

\bibitem{nufit_globalfit}
M.~Blennow, E.~Fernandez-Martinez, S.~Rosauro-Alcaraz, A.~Sousa and J.~Todd,
  ``Nufit 5.3 (2024).''.

\bibitem{Esteban:2020cvm}
I.~Esteban, M.C.~Gonzalez-Garcia, M.~Maltoni, T.~Schwetz and A.~Zhou,
  \emph{{The fate of hints: updated global analysis of three-flavor neutrino
  oscillations}}, \href{https://doi.org/10.1007/JHEP09(2020)178}{\emph{JHEP}
  {\bfseries 09} (2020) 178}
  [\href{https://arxiv.org/abs/2007.14792}{{\ttfamily 2007.14792}}].

\bibitem{Capozzi:2018ubv}
F.~Capozzi, E.~Lisi, A.~Marrone and A.~Palazzo, \emph{{Current unknowns in the
  three neutrino framework}},
  \href{https://doi.org/10.1016/j.ppnp.2018.05.005}{\emph{Prog. Part. Nucl.
  Phys.} {\bfseries 102} (2018) 48}
  [\href{https://arxiv.org/abs/1804.09678}{{\ttfamily 1804.09678}}].

\bibitem{Pontecorvo:1957qd}
B.~Pontecorvo, \emph{{Inverse beta processes and nonconservation of lepton
  charge}}, {\emph{Zh. Eksp. Teor. Fiz.} {\bfseries 34} (1957) 247}.

\bibitem{Pontecorvo:1957cp}
B.~Pontecorvo, \emph{{Mesonium and anti-mesonium}}, {\emph{Sov. Phys. JETP}
  {\bfseries 6} (1957) 429}.

\bibitem{Gribov:1968kq}
V.~Gribov and B.~Pontecorvo, \emph{{Neutrino astronomy and lepton charge}},
  \href{https://doi.org/10.1016/0370-2693(69)90525-5}{\emph{Phys. Lett. B}
  {\bfseries 28} (1969) 493}.

\bibitem{Maki:1962mu}
Z.~Maki, M.~Nakagawa and S.~Sakata, \emph{{Remarks on the unified model of
  elementary particles}}, \href{https://doi.org/10.1143/PTP.28.870}{\emph{Prog.
  Theor. Phys.} {\bfseries 28} (1962) 870}.

\bibitem{Giunti:2019aiy}
C.~Giunti and T.~Lasserre, \emph{{eV-scale Sterile Neutrinos}},
  \href{https://doi.org/10.1146/annurev-nucl-101918-023755}{\emph{Ann. Rev.
  Nucl. Part. Sci.} {\bfseries 69} (2019) 163}
  [\href{https://arxiv.org/abs/1901.08330}{{\ttfamily 1901.08330}}].

\bibitem{Dasgupta:2021ies}
B.~Dasgupta and J.~Kopp, \emph{{Sterile Neutrinos}},
  \href{https://doi.org/10.1016/j.physrep.2021.06.002}{\emph{Phys. Rept.}
  {\bfseries 928} (2021) 1} [\href{https://arxiv.org/abs/2106.05913}{{\ttfamily
  2106.05913}}].

\bibitem{Hill:1995gf}
J.E.~Hill, \emph{{Results from the LSND neutrino oscillation search for
  anti-muon-neutrino ---\ensuremath{>} anti-electron-neutrino}},
  \href{https://doi.org/10.1103/PhysRevLett.75.2654}{\emph{Phys. Rev. Lett.}
  {\bfseries 75} (1995) 2654}
  [\href{https://arxiv.org/abs/hep-ex/9504009}{{\ttfamily hep-ex/9504009}}].

\bibitem{LSND:1996ubh}
{\scshape LSND} collaboration, \emph{{Evidence for anti-muon-neutrino
  ---\ensuremath{>} anti-electron-neutrino oscillations from the LSND
  experiment at LAMPF}},
  \href{https://doi.org/10.1103/PhysRevLett.77.3082}{\emph{Phys. Rev. Lett.}
  {\bfseries 77} (1996) 3082}
  [\href{https://arxiv.org/abs/nucl-ex/9605003}{{\ttfamily nucl-ex/9605003}}].

\bibitem{LSND:1997vun}
{\scshape LSND} collaboration, \emph{{Evidence for nu(mu) ---\ensuremath{>}
  nu(e) neutrino oscillations from LSND}},
  \href{https://doi.org/10.1103/PhysRevLett.81.1774}{\emph{Phys. Rev. Lett.}
  {\bfseries 81} (1998) 1774}
  [\href{https://arxiv.org/abs/nucl-ex/9709006}{{\ttfamily nucl-ex/9709006}}].

\bibitem{LSND:2001aii}
{\scshape LSND} collaboration, \emph{{Evidence for neutrino oscillations from
  the observation of $\bar{\nu}_e$ appearance in a $\bar{\nu}_\mu$ beam}},
  \href{https://doi.org/10.1103/PhysRevD.64.112007}{\emph{Phys. Rev. D}
  {\bfseries 64} (2001) 112007}
  [\href{https://arxiv.org/abs/hep-ex/0104049}{{\ttfamily hep-ex/0104049}}].

\bibitem{MiniBooNE:2013uba}
{\scshape MiniBooNE} collaboration, \emph{{Improved Search for $\bar \nu_\mu
  \rightarrow \bar \nu_e$ Oscillations in the MiniBooNE Experiment}},
  \href{https://doi.org/10.1103/PhysRevLett.110.161801}{\emph{Phys. Rev. Lett.}
  {\bfseries 110} (2013) 161801}
  [\href{https://arxiv.org/abs/1303.2588}{{\ttfamily 1303.2588}}].

\bibitem{MiniBooNE:2020pnu}
{\scshape MiniBooNE} collaboration, \emph{{Updated MiniBooNE neutrino
  oscillation results with increased data and new background studies}},
  \href{https://doi.org/10.1103/PhysRevD.103.052002}{\emph{Phys. Rev. D}
  {\bfseries 103} (2021) 052002}
  [\href{https://arxiv.org/abs/2006.16883}{{\ttfamily 2006.16883}}].

\bibitem{Arguelles:2021meu}
C.A.~Arg\"uelles, I.~Esteban, M.~Hostert, K.J.~Kelly, J.~Kopp, P.A.N.~Machado
  et~al., \emph{{MicroBooNE and the \ensuremath{\nu}e Interpretation of the
  MiniBooNE Low-Energy Excess}},
  \href{https://doi.org/10.1103/PhysRevLett.128.241802}{\emph{Phys. Rev. Lett.}
  {\bfseries 128} (2022) 241802}
  [\href{https://arxiv.org/abs/2111.10359}{{\ttfamily 2111.10359}}].

\bibitem{Mention:2011rk}
G.~Mention, M.~Fechner, T.~Lasserre, T.A.~Mueller, D.~Lhuillier, M.~Cribier
  et~al., \emph{{The Reactor Antineutrino Anomaly}},
  \href{https://doi.org/10.1103/PhysRevD.83.073006}{\emph{Phys. Rev. D}
  {\bfseries 83} (2011) 073006}
  [\href{https://arxiv.org/abs/1101.2755}{{\ttfamily 1101.2755}}].

\bibitem{Mueller:2011nm}
T.A.~Mueller et~al., \emph{{Improved Predictions of Reactor Antineutrino
  Spectra}}, \href{https://doi.org/10.1103/PhysRevC.83.054615}{\emph{Phys. Rev.
  C} {\bfseries 83} (2011) 054615}
  [\href{https://arxiv.org/abs/1101.2663}{{\ttfamily 1101.2663}}].

\bibitem{Huber:2011wv}
P.~Huber, \emph{{On the determination of anti-neutrino spectra from nuclear
  reactors}}, \href{https://doi.org/10.1103/PhysRevC.85.029901}{\emph{Phys.
  Rev. C} {\bfseries 84} (2011) 024617}
  [\href{https://arxiv.org/abs/1106.0687}{{\ttfamily 1106.0687}}].

\bibitem{Serebrov:2020rhy}
A.~Serebrov and R.~Samoilov, \emph{{The analysis of the results of the
  Neutrino-4 experiment on search for sterile neutrino and comparison with
  results of other experiments}},
  \href{https://doi.org/10.31857/S1234567820160016}{\emph{Pisma Zh. Eksp. Teor.
  Fiz.} {\bfseries 112} (2020) 211}
  [\href{https://arxiv.org/abs/2003.03199}{{\ttfamily 2003.03199}}].

\bibitem{Serebrov:2020kmd}
A.P.~Serebrov et~al., \emph{{Search for sterile neutrinos with the Neutrino-4
  experiment and measurement results}},
  \href{https://doi.org/10.1103/PhysRevD.104.032003}{\emph{Phys. Rev. D}
  {\bfseries 104} (2021) 032003}
  [\href{https://arxiv.org/abs/2005.05301}{{\ttfamily 2005.05301}}].

\bibitem{neutrino4_nu2020}
{\scshape Neutrino-4} collaboration, A.~Serebrov, ``{Observation of sterile
  antineutrino oscillation in Neutrino-4 experiment at SM-3 reactor}.'' talk at
  Neutrino2020, June, 2020.
  "\url{https://indico.fnal.gov/event/43209/contributions/187878/attachments/129237/158638/Serebrov_Neutrino-4_25july.pdf}",
  2020.

\bibitem{Berryman:2020agd}
J.M.~Berryman and P.~Huber, \emph{{Sterile Neutrinos and the Global Reactor
  Antineutrino Dataset}},
  \href{https://doi.org/10.1007/JHEP01(2021)167}{\emph{JHEP} {\bfseries 01}
  (2021) 167} [\href{https://arxiv.org/abs/2005.01756}{{\ttfamily
  2005.01756}}].

\bibitem{Kopeikin:2021ugh}
V.~Kopeikin, M.~Skorokhvatov and O.~Titov, \emph{{Reevaluating reactor
  antineutrino spectra with new measurements of the ratio between U235 and
  Pu239 \ensuremath{\beta} spectra}},
  \href{https://doi.org/10.1103/PhysRevD.104.L071301}{\emph{Phys. Rev. D}
  {\bfseries 104} (2021) L071301}
  [\href{https://arxiv.org/abs/2103.01684}{{\ttfamily 2103.01684}}].

\bibitem{Giunti:2021kab}
C.~Giunti, Y.F.~Li, C.A.~Ternes and Z.~Xin, \emph{{Reactor antineutrino anomaly
  in light of recent flux model refinements}},
  \href{https://doi.org/10.1016/j.physletb.2022.137054}{\emph{Phys. Lett. B}
  {\bfseries 829} (2022) 137054}
  [\href{https://arxiv.org/abs/2110.06820}{{\ttfamily 2110.06820}}].

\bibitem{GALLEX:1997lja}
{\scshape GALLEX} collaboration, \emph{{Final results of the Cr-51 neutrino
  source experiments in GALLEX}},
  \href{https://doi.org/10.1016/S0370-2693(97)01562-1}{\emph{Phys. Lett. B}
  {\bfseries 420} (1998) 114}.

\bibitem{SAGE:1998fvr}
{\scshape SAGE} collaboration, \emph{{Measurement of the response of the
  Russian-American gallium experiment to neutrinos from a Cr-51 source}},
  \href{https://doi.org/10.1103/PhysRevC.59.2246}{\emph{Phys. Rev. C}
  {\bfseries 59} (1999) 2246}
  [\href{https://arxiv.org/abs/hep-ph/9803418}{{\ttfamily hep-ph/9803418}}].

\bibitem{Kaether:2010ag}
F.~Kaether, W.~Hampel, G.~Heusser, J.~Kiko and T.~Kirsten, \emph{{Reanalysis of
  the GALLEX solar neutrino flux and source experiments}},
  \href{https://doi.org/10.1016/j.physletb.2010.01.030}{\emph{Phys. Lett. B}
  {\bfseries 685} (2010) 47} [\href{https://arxiv.org/abs/1001.2731}{{\ttfamily
  1001.2731}}].

\bibitem{Giunti:2010zu}
C.~Giunti and M.~Laveder, \emph{{Statistical Significance of the Gallium
  Anomaly}}, \href{https://doi.org/10.1103/PhysRevC.83.065504}{\emph{Phys. Rev.
  C} {\bfseries 83} (2011) 065504}
  [\href{https://arxiv.org/abs/1006.3244}{{\ttfamily 1006.3244}}].

\bibitem{Barinov:2021asz}
V.V.~Barinov et~al., \emph{{Results from the Baksan Experiment on Sterile
  Transitions (BEST)}},
  \href{https://doi.org/10.1103/PhysRevLett.128.232501}{\emph{Phys. Rev. Lett.}
  {\bfseries 128} (2022) 232501}
  [\href{https://arxiv.org/abs/2109.11482}{{\ttfamily 2109.11482}}].

\bibitem{Giunti:2022btk}
C.~Giunti, Y.F.~Li, C.A.~Ternes, O.~Tyagi and Z.~Xin, \emph{{Gallium Anomaly:
  critical view from the global picture of \ensuremath{\nu}$_{e}$ and $
  {\overline{\nu}}_e $ disappearance}},
  \href{https://doi.org/10.1007/JHEP10(2022)164}{\emph{JHEP} {\bfseries 10}
  (2022) 164} [\href{https://arxiv.org/abs/2209.00916}{{\ttfamily
  2209.00916}}].

\bibitem{Giunti:2023kyo}
C.~Giunti and C.A.~Ternes, \emph{{Confronting solutions of the Gallium Anomaly
  with reactor rate data}},
  \href{https://doi.org/10.1016/j.physletb.2023.138436}{\emph{Phys. Lett. B}
  {\bfseries 849} (2024) 138436}
  [\href{https://arxiv.org/abs/2312.00565}{{\ttfamily 2312.00565}}].

\bibitem{Berryman:2021yan}
J.M.~Berryman, P.~Coloma, P.~Huber, T.~Schwetz and A.~Zhou, \emph{{Statistical
  significance of the sterile-neutrino hypothesis in the context of reactor and
  gallium data}}, \href{https://doi.org/10.1007/JHEP02(2022)055}{\emph{JHEP}
  {\bfseries 02} (2022) 055}
  [\href{https://arxiv.org/abs/2111.12530}{{\ttfamily 2111.12530}}].

\bibitem{Giunti:2022xat}
C.~Giunti, Y.F.~Li, C.A.~Ternes and Z.~Xin, \emph{{Inspection of the detection
  cross section dependence of the Gallium Anomaly}},
  \href{https://doi.org/10.1016/j.physletb.2023.137983}{\emph{Phys. Lett. B}
  {\bfseries 842} (2023) 137983}
  [\href{https://arxiv.org/abs/2212.09722}{{\ttfamily 2212.09722}}].

\bibitem{Huber:2022osv}
P.~Huber, \emph{{Testing the gallium anomaly}},
  \href{https://doi.org/10.1103/PhysRevD.107.096011}{\emph{Phys. Rev. D}
  {\bfseries 107} (2023) 096011}
  [\href{https://arxiv.org/abs/2209.02885}{{\ttfamily 2209.02885}}].

\bibitem{Brdar:2023cms}
V.~Brdar, J.~Gehrlein and J.~Kopp, \emph{{Towards resolving the gallium
  anomaly}}, \href{https://doi.org/10.1007/JHEP05(2023)143}{\emph{JHEP}
  {\bfseries 05} (2023) 143}
  [\href{https://arxiv.org/abs/2303.05528}{{\ttfamily 2303.05528}}].

\bibitem{Elliott:2023cvh}
S.R.~Elliott, V.~Gavrin and W.~Haxton, \emph{{The gallium anomaly}},
  \href{https://doi.org/10.1016/j.ppnp.2023.104082}{\emph{Prog. Part. Nucl.
  Phys.} {\bfseries 134} (2024) 104082}
  [\href{https://arxiv.org/abs/2306.03299}{{\ttfamily 2306.03299}}].

\bibitem{MicroBooNE:2021zai}
{\scshape MicroBooNE} collaboration, \emph{{Search for Neutrino-Induced
  Neutral-Current \ensuremath{\Delta} Radiative Decay in MicroBooNE and a First
  Test of the MiniBooNE Low Energy Excess under a Single-Photon Hypothesis}},
  \href{https://doi.org/10.1103/PhysRevLett.128.111801}{\emph{Phys. Rev. Lett.}
  {\bfseries 128} (2022) 111801}
  [\href{https://arxiv.org/abs/2110.00409}{{\ttfamily 2110.00409}}].

\bibitem{MicroBooNE:2021nxr}
{\scshape MicroBooNE} collaboration, \emph{{Search for an anomalous excess of
  inclusive charged-current $\nu_e$ interactions in the MicroBooNE experiment
  using Wire-Cell reconstruction}},
  \href{https://doi.org/10.1103/PhysRevD.105.112005}{\emph{Phys. Rev. D}
  {\bfseries 105} (2022) 112005}
  [\href{https://arxiv.org/abs/2110.13978}{{\ttfamily 2110.13978}}].

\bibitem{MicroBooNE:2021tya}
{\scshape MicroBooNE} collaboration, \emph{{Search for an Excess of Electron
  Neutrino Interactions in MicroBooNE Using Multiple Final-State Topologies}},
  \href{https://doi.org/10.1103/PhysRevLett.128.241801}{\emph{Phys. Rev. Lett.}
  {\bfseries 128} (2022) 241801}
  [\href{https://arxiv.org/abs/2110.14054}{{\ttfamily 2110.14054}}].

\bibitem{MicroBooNE:2021wad}
{\scshape MicroBooNE} collaboration, \emph{{Search for an anomalous excess of
  charged-current \ensuremath{\nu}e interactions without pions in the final
  state with the MicroBooNE experiment}},
  \href{https://doi.org/10.1103/PhysRevD.105.112004}{\emph{Phys. Rev. D}
  {\bfseries 105} (2022) 112004}
  [\href{https://arxiv.org/abs/2110.14065}{{\ttfamily 2110.14065}}].

\bibitem{MicroBooNE:2021pvo}
{\scshape MicroBooNE} collaboration, \emph{{Search for an anomalous excess of
  charged-current quasielastic \ensuremath{\nu}e interactions with the
  MicroBooNE experiment using Deep-Learning-based reconstruction}},
  \href{https://doi.org/10.1103/PhysRevD.105.112003}{\emph{Phys. Rev. D}
  {\bfseries 105} (2022) 112003}
  [\href{https://arxiv.org/abs/2110.14080}{{\ttfamily 2110.14080}}].

\bibitem{MicroBooNE:2022sdp}
{\scshape MicroBooNE} collaboration, \emph{{First Constraints on Light Sterile
  Neutrino Oscillations from Combined Appearance and Disappearance Searches
  with the MicroBooNE Detector}},
  \href{https://doi.org/10.1103/PhysRevLett.130.011801}{\emph{Phys. Rev. Lett.}
  {\bfseries 130} (2023) 011801}
  [\href{https://arxiv.org/abs/2210.10216}{{\ttfamily 2210.10216}}].

\bibitem{Denton:2021czb}
P.B.~Denton, \emph{{Sterile Neutrino Search with MicroBooNE\textquoteright{}s
  Electron Neutrino Disappearance Data}},
  \href{https://doi.org/10.1103/PhysRevLett.129.061801}{\emph{Phys. Rev. Lett.}
  {\bfseries 129} (2022) 061801}
  [\href{https://arxiv.org/abs/2111.05793}{{\ttfamily 2111.05793}}].

\bibitem{Zhang:2023zif}
C.~Zhang, X.~Qian and M.~Fallot, \emph{{Reactor antineutrino flux and
  anomaly}}, \href{https://doi.org/10.1016/j.ppnp.2024.104106}{\emph{Prog.
  Part. Nucl. Phys.} {\bfseries 136} (2024) 104106}
  [\href{https://arxiv.org/abs/2310.13070}{{\ttfamily 2310.13070}}].

\bibitem{PROSPECT:2020sxr}
{\scshape PROSPECT} collaboration, \emph{{Improved short-baseline neutrino
  oscillation search and energy spectrum measurement with the PROSPECT
  experiment at HFIR}},
  \href{https://doi.org/10.1103/PhysRevD.103.032001}{\emph{Phys. Rev. D}
  {\bfseries 103} (2021) 032001}
  [\href{https://arxiv.org/abs/2006.11210}{{\ttfamily 2006.11210}}].

\bibitem{STEREO:2022nzk}
{\scshape STEREO} collaboration, \emph{{STEREO neutrino spectrum of $^{235}$U
  fission rejects sterile neutrino hypothesis}},
  \href{https://doi.org/10.1038/s41586-022-05568-2}{\emph{Nature} {\bfseries
  613} (2023) 257} [\href{https://arxiv.org/abs/2210.07664}{{\ttfamily
  2210.07664}}].

\bibitem{DANSS:2018fnn}
{\scshape DANSS} collaboration, \emph{{Search for sterile neutrinos at the
  DANSS experiment}},
  \href{https://doi.org/10.1016/j.physletb.2018.10.038}{\emph{Phys. Lett. B}
  {\bfseries 787} (2018) 56}
  [\href{https://arxiv.org/abs/1804.04046}{{\ttfamily 1804.04046}}].

\bibitem{NEOS:2016wee}
{\scshape NEOS} collaboration, \emph{{Sterile Neutrino Search at the NEOS
  Experiment}},
  \href{https://doi.org/10.1103/PhysRevLett.118.121802}{\emph{Phys. Rev. Lett.}
  {\bfseries 118} (2017) 121802}
  [\href{https://arxiv.org/abs/1610.05134}{{\ttfamily 1610.05134}}].

\bibitem{RENO:2020hva}
{\scshape RENO, NEOS} collaboration, \emph{{Search for sterile neutrino
  oscillations using RENO and NEOS data}},
  \href{https://doi.org/10.1103/PhysRevD.105.L111101}{\emph{Phys. Rev. D}
  {\bfseries 105} (2022) L111101}
  [\href{https://arxiv.org/abs/2011.00896}{{\ttfamily 2011.00896}}].

\bibitem{Dentler:2018sju}
M.~Dentler, A.~Hern\'andez-Cabezudo, J.~Kopp, P.A.N.~Machado, M.~Maltoni,
  I.~Martinez-Soler et~al., \emph{{Updated Global Analysis of Neutrino
  Oscillations in the Presence of eV-Scale Sterile Neutrinos}},
  \href{https://doi.org/10.1007/JHEP08(2018)010}{\emph{JHEP} {\bfseries 08}
  (2018) 010} [\href{https://arxiv.org/abs/1803.10661}{{\ttfamily
  1803.10661}}].

\bibitem{Diaz:2019fwt}
A.~Diaz, C.A.~Arg\"uelles, G.H.~Collin, J.M.~Conrad and M.H.~Shaevitz,
  \emph{{Where Are We With Light Sterile Neutrinos?}},
  \href{https://doi.org/10.1016/j.physrep.2020.08.005}{\emph{Phys. Rept.}
  {\bfseries 884} (2020) 1} [\href{https://arxiv.org/abs/1906.00045}{{\ttfamily
  1906.00045}}].

\bibitem{Hardin:2022muu}
J.M.~Hardin, I.~Martinez-Soler, A.~Diaz, M.~Jin, N.W.~Kamp, C.A.~Arg\"uelles
  et~al., \emph{{New Clues about light sterile neutrinos: preference for models
  with damping effects in global fits}},
  \href{https://doi.org/10.1007/JHEP09(2023)058}{\emph{JHEP} {\bfseries 09}
  (2023) 058} [\href{https://arxiv.org/abs/2211.02610}{{\ttfamily
  2211.02610}}].

\bibitem{Acciarri:2015uup}
{\scshape DUNE} collaboration, \emph{{Long-Baseline Neutrino Facility (LBNF)
  and Deep Underground Neutrino Experiment (DUNE) Conceptual Design Report
  Volume 2: The Physics Program for DUNE at LBNF}},
  \href{https://arxiv.org/abs/1512.06148}{{\ttfamily 1512.06148}}.

\bibitem{DUNE:2020ypp}
{\scshape DUNE} collaboration, \emph{{Deep Underground Neutrino Experiment
  (DUNE), Far Detector Technical Design Report, Volume II: DUNE Physics}},
  \href{https://arxiv.org/abs/2002.03005}{{\ttfamily 2002.03005}}.

\bibitem{DUNE:2021cuw}
{\scshape DUNE} collaboration, \emph{{Experiment Simulation Configurations
  Approximating DUNE TDR}},  \href{https://arxiv.org/abs/2103.04797}{{\ttfamily
  2103.04797}}.

\bibitem{DUNE:2020fgq}
{\scshape DUNE} collaboration, \emph{{Prospects for beyond the Standard Model
  physics searches at the Deep Underground Neutrino Experiment}},
  \href{https://doi.org/10.1140/epjc/s10052-021-09007-w}{\emph{Eur. Phys. J. C}
  {\bfseries 81} (2021) 322}
  [\href{https://arxiv.org/abs/2008.12769}{{\ttfamily 2008.12769}}].

\bibitem{Hyper-KamiokandeProto-:2015xww}
{\scshape Hyper-Kamiokande Proto-} collaboration, \emph{{Physics potential of a
  long-baseline neutrino oscillation experiment using a J-PARC neutrino beam
  and Hyper-Kamiokande}},
  \href{https://doi.org/10.1093/ptep/ptv061}{\emph{PTEP} {\bfseries 2015}
  (2015) 053C02} [\href{https://arxiv.org/abs/1502.05199}{{\ttfamily
  1502.05199}}].

\bibitem{Hyper-Kamiokande:2016srs}
{\scshape Hyper-Kamiokande} collaboration, \emph{{Physics potentials with the
  second Hyper-Kamiokande detector in Korea}},
  \href{https://doi.org/10.1093/ptep/pty044}{\emph{PTEP} {\bfseries 2018}
  (2018) 063C01} [\href{https://arxiv.org/abs/1611.06118}{{\ttfamily
  1611.06118}}].

\bibitem{ESSnuSB:2013dql}
{\scshape ESSnuSB} collaboration, \emph{{A very intense neutrino super beam
  experiment for leptonic CP violation discovery based on the European
  spallation source linac}},
  \href{https://doi.org/10.1016/j.nuclphysb.2014.05.016}{\emph{Nucl. Phys. B}
  {\bfseries 885} (2014) 127}
  [\href{https://arxiv.org/abs/1309.7022}{{\ttfamily 1309.7022}}].

\bibitem{Berryman:2015nua}
J.M.~Berryman, A.~de~Gouv\^ea, K.J.~Kelly and A.~Kobach, \emph{{Sterile
  neutrino at the Deep Underground Neutrino Experiment}},
  \href{https://doi.org/10.1103/PhysRevD.92.073012}{\emph{Phys. Rev. D}
  {\bfseries 92} (2015) 073012}
  [\href{https://arxiv.org/abs/1507.03986}{{\ttfamily 1507.03986}}].

\bibitem{Gandhi:2015xza}
R.~Gandhi, B.~Kayser, M.~Masud and S.~Prakash, \emph{{The impact of sterile
  neutrinos on CP measurements at long baselines}},
  \href{https://doi.org/10.1007/JHEP11(2015)039}{\emph{JHEP} {\bfseries 11}
  (2015) 039} [\href{https://arxiv.org/abs/1508.06275}{{\ttfamily
  1508.06275}}].

\bibitem{Agarwalla:2016xxa}
S.K.~Agarwalla, S.S.~Chatterjee and A.~Palazzo, \emph{{Physics Reach of DUNE
  with a Light Sterile Neutrino}},
  \href{https://doi.org/10.1007/JHEP09(2016)016}{\emph{JHEP} {\bfseries 09}
  (2016) 016} [\href{https://arxiv.org/abs/1603.03759}{{\ttfamily
  1603.03759}}].

\bibitem{Agarwalla:2016xlg}
S.K.~Agarwalla, S.S.~Chatterjee and A.~Palazzo, \emph{{Octant of $\theta_{23}$
  in danger with a light sterile neutrino}},
  \href{https://doi.org/10.1103/PhysRevLett.118.031804}{\emph{Phys. Rev. Lett.}
  {\bfseries 118} (2017) 031804}
  [\href{https://arxiv.org/abs/1605.04299}{{\ttfamily 1605.04299}}].

\bibitem{Dutta:2016glq}
D.~Dutta, R.~Gandhi, B.~Kayser, M.~Masud and S.~Prakash, \emph{{Capabilities of
  long-baseline experiments in the presence of a sterile neutrino}},
  \href{https://doi.org/10.1007/JHEP11(2016)122}{\emph{JHEP} {\bfseries 11}
  (2016) 122} [\href{https://arxiv.org/abs/1607.02152}{{\ttfamily
  1607.02152}}].

\bibitem{Blennow:2016jkn}
M.~Blennow, P.~Coloma, E.~Fernandez-Martinez, J.~Hernandez-Garcia and
  J.~Lopez-Pavon, \emph{{Non-Unitarity, sterile neutrinos, and Non-Standard
  neutrino Interactions}},
  \href{https://doi.org/10.1007/JHEP04(2017)153}{\emph{JHEP} {\bfseries 04}
  (2017) 153} [\href{https://arxiv.org/abs/1609.08637}{{\ttfamily
  1609.08637}}].

\bibitem{Rout:2017udo}
J.~Rout, M.~Masud and P.~Mehta, \emph{{Can we probe intrinsic CP and T
  violations and nonunitarity at long baseline accelerator experiments?}},
  \href{https://doi.org/10.1103/PhysRevD.95.075035}{\emph{Phys. Rev. D}
  {\bfseries 95} (2017) 075035}
  [\href{https://arxiv.org/abs/1702.02163}{{\ttfamily 1702.02163}}].

\bibitem{Choubey:2017cba}
S.~Choubey, D.~Dutta and D.~Pramanik, \emph{{Imprints of a light Sterile
  Neutrino at DUNE, T2HK and T2HKK}},
  \href{https://doi.org/10.1103/PhysRevD.96.056026}{\emph{Phys. Rev. D}
  {\bfseries 96} (2017) 056026}
  [\href{https://arxiv.org/abs/1704.07269}{{\ttfamily 1704.07269}}].

\bibitem{Coloma:2017ptb}
P.~Coloma, D.V.~Forero and S.J.~Parke, \emph{{DUNE Sensitivities to the Mixing
  between Sterile and Tau Neutrinos}},
  \href{https://doi.org/10.1007/JHEP07(2018)079}{\emph{JHEP} {\bfseries 07}
  (2018) 079} [\href{https://arxiv.org/abs/1707.05348}{{\ttfamily
  1707.05348}}].

\bibitem{Gandhi:2017vzo}
R.~Gandhi, B.~Kayser, S.~Prakash and S.~Roy, \emph{{What measurements of
  neutrino neutral current events can reveal}},
  \href{https://doi.org/10.1007/JHEP11(2017)202}{\emph{JHEP} {\bfseries 11}
  (2017) 202} [\href{https://arxiv.org/abs/1708.01816}{{\ttfamily
  1708.01816}}].

\bibitem{Tang:2017khg}
J.~Tang, Y.~Zhang and Y.-F.~Li, \emph{{Probing Direct and Indirect Unitarity
  Violation in Future Accelerator Neutrino Facilities}},
  \href{https://doi.org/10.1016/j.physletb.2017.09.055}{\emph{Phys. Lett. B}
  {\bfseries 774} (2017) 217}
  [\href{https://arxiv.org/abs/1708.04909}{{\ttfamily 1708.04909}}].

\bibitem{Choubey:2017ppj}
S.~Choubey, D.~Dutta and D.~Pramanik, \emph{{Measuring the Sterile Neutrino CP
  Phase at DUNE and T2HK}},
  \href{https://doi.org/10.1140/epjc/s10052-018-5816-y}{\emph{Eur. Phys. J. C}
  {\bfseries 78} (2018) 339}
  [\href{https://arxiv.org/abs/1711.07464}{{\ttfamily 1711.07464}}].

\bibitem{Chatla:2018sos}
A.~Chatla, S.~Rudrabhatla and B.A.~Bambah, \emph{{Degeneracy Resolution
  Capabilities of NO$\nu$A and DUNE in the Presence of Light Sterile
  Neutrino}}, \href{https://doi.org/10.1155/2018/2547358}{\emph{Adv. High
  Energy Phys.} {\bfseries 2018} (2018) 2547358}
  [\href{https://arxiv.org/abs/1804.02818}{{\ttfamily 1804.02818}}].

\bibitem{Choubey:2018kqq}
S.~Choubey, D.~Dutta and D.~Pramanik, \emph{{Exploring fake solutions in the
  sterile neutrino sector at long-baseline experiments}},
  \href{https://doi.org/10.1140/epjc/s10052-019-7479-8}{\emph{Eur. Phys. J. C}
  {\bfseries 79} (2019) 968}
  [\href{https://arxiv.org/abs/1811.08684}{{\ttfamily 1811.08684}}].

\bibitem{Ghoshal:2019pab}
A.~Ghoshal, A.~Giarnetti and D.~Meloni, \emph{{On the role of the $\nu_{?}$
  appearance in DUNE in constraining standard neutrino physics and beyond}},
  \href{https://doi.org/10.1007/JHEP12(2019)126}{\emph{JHEP} {\bfseries 12}
  (2019) 126} [\href{https://arxiv.org/abs/1906.06212}{{\ttfamily
  1906.06212}}].

\bibitem{Krasnov:2019kdc}
I.~Krasnov, \emph{{DUNE prospects in the search for sterile neutrinos}},
  \href{https://doi.org/10.1103/PhysRevD.100.075023}{\emph{Phys. Rev. D}
  {\bfseries 100} (2019) 075023}
  [\href{https://arxiv.org/abs/1902.06099}{{\ttfamily 1902.06099}}].

\bibitem{Majhi:2019hdj}
R.~Majhi, C.~Soumya and R.~Mohanta, \emph{{Light sterile neutrinos and their
  implications on currently running long-baseline and neutrinoless double beta
  decay experiments}}, \href{https://doi.org/10.1088/1361-6471/ab9797}{\emph{J.
  Phys. G} {\bfseries 47} (2020) 095002}
  [\href{https://arxiv.org/abs/1911.10952}{{\ttfamily 1911.10952}}].

\bibitem{Fiza:2021gvq}
N.~Fiza, M.~Masud and M.~Mitra, \emph{{Exploring the new physics phases in 3+1
  scenario in neutrino oscillation experiments}},
  \href{https://doi.org/10.1007/JHEP09(2021)162}{\emph{JHEP} {\bfseries 09}
  (2021) 162} [\href{https://arxiv.org/abs/2102.05063}{{\ttfamily
  2102.05063}}].

\bibitem{Ghosh:2021rtn}
M.~Ghosh and R.~Mohanta, \emph{{Updated sensitivity of DUNE in 3 + 1 scenario
  with far and near detectors}},
  \href{https://doi.org/10.1140/epjs/s11734-021-00360-1}{\emph{Eur. Phys. J.
  ST} {\bfseries 231} (2022) 137}
  [\href{https://arxiv.org/abs/2110.05767}{{\ttfamily 2110.05767}}].

\bibitem{Penedo:2022etl}
J.T.~Penedo and J.a.~Pulido, \emph{{Baseline and other effects for a sterile
  neutrino at DUNE}},
  \href{https://doi.org/10.1103/PhysRevD.107.075026}{\emph{Phys. Rev. D}
  {\bfseries 107} (2023) 075026}
  [\href{https://arxiv.org/abs/2207.02331}{{\ttfamily 2207.02331}}].

\bibitem{Denton:2022pxt}
P.B.~Denton, A.~Giarnetti and D.~Meloni, \emph{{How to identify different new
  neutrino oscillation physics scenarios at DUNE}},
  \href{https://doi.org/10.1007/JHEP02(2023)210}{\emph{JHEP} {\bfseries 02}
  (2023) 210} [\href{https://arxiv.org/abs/2210.00109}{{\ttfamily
  2210.00109}}].

\bibitem{Chatterjee:2022pqg}
A.~Chatterjee, S.~Goswami and S.~Pan, \emph{{Matter effect in presence of a
  sterile neutrino and resolution of the octant degeneracy using a liquid argon
  detector}}, \href{https://doi.org/10.1103/PhysRevD.108.095050}{\emph{Phys.
  Rev. D} {\bfseries 108} (2023) 095050}
  [\href{https://arxiv.org/abs/2212.02949}{{\ttfamily 2212.02949}}].

\bibitem{Parveen:2023ixk}
S.~Parveen, K.~Sharma, S.~Patra and P.~Mehta, \emph{{$CP$ and $T$ violation
  effects in presence of an eV scale sterile neutrino at long baseline neutrino
  experiments}},  \href{https://arxiv.org/abs/2305.16824}{{\ttfamily
  2305.16824}}.

\bibitem{Kaur:2024jko}
D.~Kaur, \emph{{\ensuremath{\theta}23 Octant sensitivity in presence of light
  sterile and active \ensuremath{\nu} and \ensuremath{\bar{\nu}} oscillations
  using beamline experiments}},
  \href{https://doi.org/10.1016/j.nuclphysb.2024.116517}{\emph{Nucl. Phys. B}
  {\bfseries 1002} (2024) 116517}.

\bibitem{Farzan:2017xzy}
Y.~Farzan and M.~Tortola, \emph{{Neutrino oscillations and Non-Standard
  Interactions}}, \href{https://doi.org/10.3389/fphy.2018.00010}{\emph{Front.
  in Phys.} {\bfseries 6} (2018) 10}
  [\href{https://arxiv.org/abs/1710.09360}{{\ttfamily 1710.09360}}].

\bibitem{Proceedings:2019qno}
\emph{{Neutrino Non-Standard Interactions: A Status Report}}, vol.~2, 2019.
\newblock 10.21468/SciPostPhysProc.2.001.

\bibitem{dunefluxes}
DUNE Fluxes, https://glaucus.crc.nd.edu/DUNEFluxes/.

\bibitem{Masud:2017bcf}
M.~Masud, M.~Bishai and P.~Mehta, \emph{{Extricating New Physics Scenarios at
  DUNE with Higher Energy Beams}},
  \href{https://doi.org/10.1038/s41598-018-36790-6}{\emph{Sci. Rep.} {\bfseries
  9} (2019) 352} [\href{https://arxiv.org/abs/1704.08650}{{\ttfamily
  1704.08650}}].

\bibitem{Masud:2018pig}
M.~Masud, S.~Roy and P.~Mehta, \emph{{Correlations and degeneracies among the
  NSI parameters with tunable beams at DUNE}},
  \href{https://doi.org/10.1103/PhysRevD.99.115032}{\emph{Phys. Rev. D}
  {\bfseries 99} (2019) 115032}
  [\href{https://arxiv.org/abs/1812.10290}{{\ttfamily 1812.10290}}].

\bibitem{Rout:2020cxi}
J.~Rout, S.~Roy, M.~Masud, M.~Bishai and P.~Mehta, \emph{{Impact of high energy
  beam tunes on the sensitivities to the standard unknowns at DUNE}},
  \href{https://doi.org/10.1103/PhysRevD.102.116018}{\emph{Phys. Rev. D}
  {\bfseries 102} (2020) 116018}
  [\href{https://arxiv.org/abs/2009.05061}{{\ttfamily 2009.05061}}].

\bibitem{Rout:2020emr}
J.~Rout, S.~Shafaq, M.~Bishai and P.~Mehta, \emph{{Physics prospects with the
  second oscillation maximum at the Deep Underground Neutrino Experiment}},
  \href{https://doi.org/10.1103/PhysRevD.103.116003}{\emph{Phys. Rev. D}
  {\bfseries 103} (2021) 116003}
  [\href{https://arxiv.org/abs/2012.08269}{{\ttfamily 2012.08269}}].

\bibitem{Siyeon:2024pte}
K.~Siyeon, S.~Kim, M.~Masud and J.~Park, \emph{{Probing large extra dimension
  at DUNE using beam tunes}},
  \href{https://doi.org/10.1007/JHEP11(2024)141}{\emph{JHEP} {\bfseries 11}
  (2024) 141} [\href{https://arxiv.org/abs/2409.08620}{{\ttfamily
  2409.08620}}].

\bibitem{Kopp:2013vaa}
J.~Kopp, P.A.N.~Machado, M.~Maltoni and T.~Schwetz, \emph{{Sterile Neutrino
  Oscillations: The Global Picture}},
  \href{https://doi.org/10.1007/JHEP05(2013)050}{\emph{JHEP} {\bfseries 05}
  (2013) 050} [\href{https://arxiv.org/abs/1303.3011}{{\ttfamily 1303.3011}}].

\bibitem{Klop:2014ima}
N.~Klop and A.~Palazzo, \emph{{Imprints of CP violation induced by sterile
  neutrinos in T2K data}},
  \href{https://doi.org/10.1103/PhysRevD.91.073017}{\emph{Phys. Rev. D}
  {\bfseries 91} (2015) 073017}
  [\href{https://arxiv.org/abs/1412.7524}{{\ttfamily 1412.7524}}].

\bibitem{:2019wbn}
Y.~Reyimuaji and C.~Liu, \emph{{Prospects of light sterile neutrino searches in
  long-baseline neutrino oscillations}},
  \href{https://doi.org/10.1007/JHEP06(2020)094}{\emph{JHEP} {\bfseries 06}
  (2020) 094} [\href{https://arxiv.org/abs/1911.12524}{{\ttfamily
  1911.12524}}].

\bibitem{Huber:2004ka}
P.~Huber, M.~Lindner and W.~Winter, \emph{{Simulation of long-baseline neutrino
  oscillation experiments with GLoBES (General Long Baseline Experiment
  Simulator)}}, \href{https://doi.org/10.1016/j.cpc.2005.01.003}{\emph{Comput.
  Phys. Commun.} {\bfseries 167} (2005) 195}
  [\href{https://arxiv.org/abs/hep-ph/0407333}{{\ttfamily hep-ph/0407333}}].

\bibitem{Huber:2007ji}
P.~Huber, J.~Kopp, M.~Lindner, M.~Rolinec and W.~Winter, \emph{{New features in
  the simulation of neutrino oscillation experiments with GLoBES 3.0: General
  Long Baseline Experiment Simulator}},
  \href{https://doi.org/10.1016/j.cpc.2007.05.004}{\emph{Comput. Phys. Commun.}
  {\bfseries 177} (2007) 432}
  [\href{https://arxiv.org/abs/hep-ph/0701187}{{\ttfamily hep-ph/0701187}}].

\bibitem{Gandhi:2004bj}
R.~Gandhi, P.~Ghoshal, S.~Goswami, P.~Mehta and S.U.~Sankar, \emph{{Earth
  matter effects at very long baselines and the neutrino mass hierarchy}},
  \href{https://doi.org/10.1103/PhysRevD.73.053001}{\emph{Phys. Rev. D}
  {\bfseries 73} (2006) 053001}
  [\href{https://arxiv.org/abs/hep-ph/0411252}{{\ttfamily hep-ph/0411252}}].

\bibitem{Agostinelli:2002hh}
{\scshape GEANT4} collaboration, \emph{{GEANT4: A Simulation toolkit}},
  \href{https://doi.org/10.1016/S0168-9002(03)01368-8}{\emph{Nucl. Instrum.
  Meth.} {\bfseries A506} (2003) 250}.

\bibitem{Allison:2006ve}
J.~Allison et~al., \emph{{Geant4 developments and applications}},
  \href{https://doi.org/10.1109/TNS.2006.869826}{\emph{IEEE Trans. Nucl. Sci.}
  {\bfseries 53} (2006) 270}.

\bibitem{Huber:2002mx}
P.~Huber, M.~Lindner and W.~Winter, \emph{{Superbeams versus neutrino
  factories}}, \href{https://doi.org/10.1016/S0550-3213(02)00825-8}{\emph{Nucl.
  Phys.} {\bfseries B645} (2002) 3}
  [\href{https://arxiv.org/abs/hep-ph/0204352}{{\ttfamily hep-ph/0204352}}].

\bibitem{Fogli:2002pt}
G.L.~Fogli, E.~Lisi, A.~Marrone, D.~Montanino and A.~Palazzo, \emph{{Getting
  the most from the statistical analysis of solar neutrino oscillations}},
  \href{https://doi.org/10.1103/PhysRevD.66.053010}{\emph{Phys. Rev.}
  {\bfseries D66} (2002) 053010}
  [\href{https://arxiv.org/abs/hep-ph/0206162}{{\ttfamily hep-ph/0206162}}].

\bibitem{GonzalezGarcia:2004wg}
M.~Gonzalez-Garcia and M.~Maltoni, \emph{{Atmospheric neutrino oscillations and
  new physics}},
  \href{https://doi.org/10.1103/PhysRevD.70.033010}{\emph{Phys.Rev.} {\bfseries
  D70} (2004) 033010} [\href{https://arxiv.org/abs/hep-ph/0404085}{{\ttfamily
  hep-ph/0404085}}].

\bibitem{Gandhi:2007td}
R.~Gandhi, P.~Ghoshal, S.~Goswami, P.~Mehta, S.U.~Sankar and S.~Shalgar,
  \emph{{Mass Hierarchy Determination via future Atmospheric Neutrino
  Detectors}}, \href{https://doi.org/10.1103/PhysRevD.76.073012}{\emph{Phys.
  Rev.} {\bfseries D76} (2007) 073012}
  [\href{https://arxiv.org/abs/0707.1723}{{\ttfamily 0707.1723}}].

\bibitem{Qian:2012zn}
X.~Qian, A.~Tan, W.~Wang, J.J.~Ling, R.D.~McKeown and C.~Zhang,
  \emph{{Statistical Evaluation of Experimental Determinations of Neutrino Mass
  Hierarchy}}, \href{https://doi.org/10.1103/PhysRevD.86.113011}{\emph{Phys.
  Rev.} {\bfseries D86} (2012) 113011}
  [\href{https://arxiv.org/abs/1210.3651}{{\ttfamily 1210.3651}}].

\bibitem{ParticleDataGroup:2022pth}
{\scshape Particle Data Group} collaboration, \emph{{Review of Particle
  Physics}}, \href{https://doi.org/10.1093/ptep/ptac097}{\emph{PTEP} {\bfseries
  2022} (2022) 083C01}.

\bibitem{Singha:2022btw}
D.K.~Singha, M.~Ghosh, R.~Majhi and R.~Mohanta, \emph{{Study of light sterile
  neutrino at the long-baseline experiment options at KM3NeT}},
  \href{https://doi.org/10.1103/PhysRevD.107.075039}{\emph{Phys. Rev. D}
  {\bfseries 107} (2023) 075039}
  [\href{https://arxiv.org/abs/2211.01816}{{\ttfamily 2211.01816}}].

\bibitem{DayaBay:2016lkk}
{\scshape Daya Bay, MINOS} collaboration, \emph{{Limits on Active to Sterile
  Neutrino Oscillations from Disappearance Searches in the MINOS, Daya Bay, and
  Bugey-3 Experiments}},
  \href{https://doi.org/10.1103/PhysRevLett.117.151801}{\emph{Phys. Rev. Lett.}
  {\bfseries 117} (2016) 151801}
  [\href{https://arxiv.org/abs/1607.01177}{{\ttfamily 1607.01177}}].

\bibitem{MINOS:2017cae}
{\scshape MINOS+} collaboration, \emph{{Search for sterile neutrinos in MINOS
  and MINOS+ using a two-detector fit}},
  \href{https://doi.org/10.1103/PhysRevLett.122.091803}{\emph{Phys. Rev. Lett.}
  {\bfseries 122} (2019) 091803}
  [\href{https://arxiv.org/abs/1710.06488}{{\ttfamily 1710.06488}}].

\end{thebibliography}\endgroup
\end{document}